\documentclass[epj]{svjour}
\usepackage{graphics}
\usepackage{graphicx}
\usepackage{amsmath}
\usepackage{hyperref}
\usepackage{amssymb}
\usepackage{subfigure}
\usepackage{epic}
\usepackage{eepic}
\newcounter{subequation}
 \renewcommand{\theequation}{\arabic{section}.\arabic{equation}\ifnum\thesubequation>0{\alph{subequation}}\fi}

 \def\c{{\mathbf{c}}}
 \def\dis{{\rm dis}}
 \def\anis{{\rm anis}}
 \def\e{{\mathbf{e}}}
 \def\p{{\mathbf{p}}}
 \def\R{{\mathbb{R}}}
 
 \def\x{{\mathbf{x}}}
 \def\V{{\mathbf{V}}}
 \def\n{{\mathbf{n}}}
 \def\d{{\rm{d}}}
 \def\tr{{\mathbf{Tr}}}

\begin{document}

\title{Extended  morphometric analysis    of   neuronal   cells   with
  Minkowski   valuations}

\author{Claus Beisbart\inst{1,2}, Marconi S.  Barbosa\inst{3}, Herbert
Wagner\inst{2} \and Luciano da F. Costa\inst{3} }

 \institute{
P.P.M. Research Group\\
Center for Junior Research Fellows, University of Konstanz, P.O. Box M 682\\ D--78457 Konstanz, Germany
\and  
Arnold Sommerfeld Center for Theoretical Physics, Ludwig-Maximilians-Universit\"at \\
             Theresienstra\ss e 37 \\
            D--80333 M\"unchen, 
            Germany 
            \and     
            Cybernetic Vision Research Group, DFI-IFSC\\ 
            Universidade de S\~ ao Paulo, S\~{a}o Carlos, SP\\ 
            Caixa Postal 369, 13560-970, Brasil
            }

\date{Received: date / Revised version: date}

\abstract{Minkowski  valuations  provide  a systematic  framework  for
  quantifying different aspects of  morphology. In this paper we apply
  vector-  and tensor-valued  Minkowski valuations  to  neuronal cells
  from  the cat's  retina  in order  to  describe their  morphological
  structure  in a  comprehensive way.  We introduce  the  framework of
  Minkowski  valuations,  discuss  their implementation  for  neuronal
  cells and show how they  can discriminate between cells of different
  types.     \PACS{   {07.05.Kf}{Data    analysis:    algorithms   and
  implementation; data  management} \and {87.19.La}{Neuroscience} \and
  {02.40.Ft}{Convex sets  and geometric inequalities} } %  end of PACS
  codes }

  \authorrunning{Beisbart,   Barbosa,   Wagner   \&   da   F.   Costa}
  \titlerunning{Minkowski valuations for neuronal cells} \maketitle

\section{Introduction}

Natural phenomena  can be understood  as causes and consequences  of a
continuing  interplay  between  geometry  and dynamics,  or  form  and
function~\cite{Douady&Coulder:1992,Percolation:2003}.           Spatial
adjacencies, the specific geometrical features of natural entities, as
well  as the  dimensionality of  the  space where  they are  embedded,
largely  constrain their  dynamics  and function.   For instance,  the
proper operation of  a mammal's heart depends on  a suitable diffusion
of  potentials and  waves  across the  heart  surface.  It  is at  the
central nervous  system, however, that the interplay  between form and
function reaches its greatest complexity.  To begin with, the velocity
of signal  transmission in  neuronal fibers (i.e.  dendrites and axons) depends  on the  width and
length of  the fibers.  The  importance of
geometry  for proper neuronal  function is  further underlined  by the
fact  that  neurons  are  cells  specialized  to  establish  selective
connections.     Given   the    spatial    constraints   imposed    by
three-dimensional  space,  these cells  have  to  resort  to the  most
diverse geometries  in order to implement  the required interconnections
-- as they do  in a dynamical way during the  whole lifetime of an
individual.
\\
As a consequence  it is to be expected  that morphological analyses of
    neuronal cells  provide clues for  understanding neural dynamics
    and function.  Although a large number  of investigations have
    been   directed    at   the   neural    anatomy   and   geometry
    (e.g.~\cite{wassle:1981,fuckuda:1984,wassle:1981,pelt:2002}), only
    the taxonomic organization of  neuronal cells or the consideration
    of  shape  abnormalities  as  subsidies for  diagnosis  have  been
    concentrated on so far.   The study of  the shape of neuronal  cells with
    objective  and   mathematically  well  characterized  morphometric
    descriptors       is       just       at       the       beginning
    (e.g.~\cite{velte:1999,Ascoli_Krichmar:2000,Coelho_Costa:2001,Potts_PNAS:2003}).
\\
In order to be useful tools, such morphological descriptors should fulfill the
following criteria:  First, the  extracted  quantitative features  should
obey simple transformation  rules, when the neuronal cell under investigation is subject to
elementary  geometrical  transformations   such  as  affine  or  conformal
transformations  ({\em  in-}  and  {\em covariance}).   Second, the  obtained
measurements should  discriminate between different classes of
 neuronal cells.  Finally,  it is important that the estimated
features  allow for  intuitive interpretations  from  the neuroscience
point of view.
\\
Because of their long tradition in modeling and analysis, mathematics,
physics  and  engineering  provide  a  large number  of  concepts  and
measures  that may  be considered  for studies  in  neuroscience and
neuromorphometry.  A good  example is entropy, which has  been used in
neuroscience  because of  its close  association with  the  concept of
information~\cite{Rieke:1999}.    Other  such  measures   include  the
fractal         dimension~\cite{Morigiwa:1989,Coelho:1996,Manoel:2002},
lacunarity~\cite{Smith:1996,Rodrigues:2005},    percolation   critical
density~\cite{Percolation:2003}    and    curvature~\cite{Cesar:1998}.
Recently,  concepts  from Integral  Geometry  and  in particular  the
(scalar)  Minkowski  shape  functionals   were  applied  in  order  to
characterize   the  geometry   of  ganglion   cells  from   the  cat's
re\-tina~\cite{Barbosa:2003a,Barbosa:2003b}. Minkowski shape functionals
are particularly interesting because  they meet the criteria mentioned
above: They  are invariant under  rigid body transformations,  seem to
have  good  discriminative   power~\cite{Barbosa:2003a},  and  can  be
squared  with basic  concepts  from neuroscience.  Moreover, they  can
easily  be implemented:   Usually,  the  original data  are filtered  with   methods  known  from   MIA  (Morphological  Image
Analysis). This  preprocessing introduces a free parameter,  which can later
be  varied in order  to probe the morphology at different scales. In
previous  works,  the   singular  points  (branching  and  terminating
points)~\cite{Barbosa:2003b}       or       the       whole       cell
outline~\cite{Barbosa:2003a} were  dilated, where the  dilation radius
entered  as   a    parameter.   For  each   dilation  radius,  the
preprocessed  neuron image  was decomposed  into components  (or basic
building blocks). The  Minkowski functionals can then  be calculated by
counting certain  multiplicities of  the basic building  blocks.  This
approach   makes   use   of   mathematical   results   from   Integral
Geometry~\cite{Raedt:2001}.
\\
In this paper,  we use extensions of the  Minkowski shape functionals,
viz.  the  {\em Min\-kowski valuations},  in order to  further improve
neuromorphometric  characterization  and  ana\-lysis.  These  extensions
were       only        very       recently       investigated       by
mathematicians~\cite{alesker:tensor,Schneider:2000,Schneider:2002} and
include  vector-  and   ten\-sor-valued  measures.  They  are  therefore
sensitive  to  directional information  and  also  allow for  valuable
graphical visualizations.  Minkowski valuations have been successfully
applied to  describe the morphology  of galaxies~\cite{Claus:2002} and
galaxy clusters~\cite{Claus:2001}.
\\
In the following, we will illustrate the potential of the Minkowski valuations
for neuromorphometry by analyzing a set of ten ganglion cells from the
cat's retina.  We consider two-dimensional projections of the
cells. The set used~\cite{Masland:2001} includes cells with diverse
shapes, corresponding  to a  recently revised classification  of those
types of  cells~\cite{Berson:2002}.  In addition,  ganglion cells from
the retina exhibit branching  patterns which are predominantly planar,
and therefore compatible with the two-dimensional Minkowski valuations
considered in the present work.
\\
The   article  starts   by  presenting   the   higher-order  Min\-kows\-ki
functionals  and proceeds  by  illustrating their  application to  the
characterization of neuronal cells.

%%%%%%%%%%%%%%%%%%%%%%%%%%%%%%%%%%%%%%%%%%%%%%%%%%
\section{Minkowski valuations}
\begin{figure}
\centering\includegraphics[width=8.7cm]{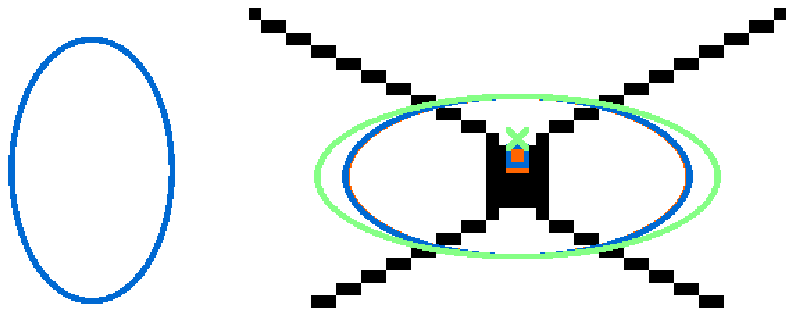}
\centering\includegraphics[width=8.7cm]{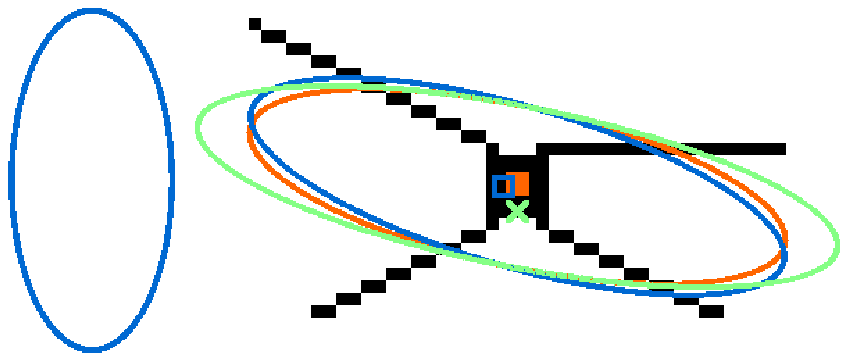}
\centering\includegraphics[width=8.7cm]{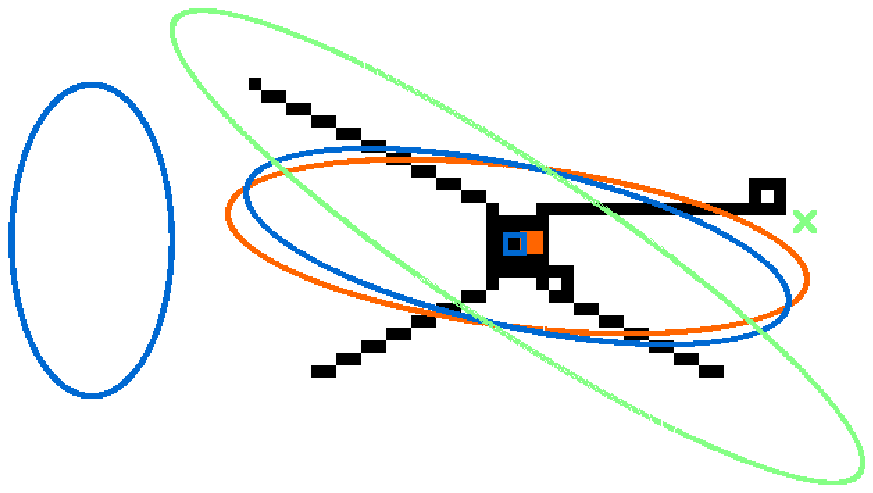}
\caption{Three toy  examples to be  discussed as an  illustration. For
  the  centroids  the following  point  styles  are  used: red (medium  grey)
  filled square:  $\p_0$;  blue (dark grey)  open  square:  $\p_1$;
  green (light grey)  x:
  $\p_2$.  The ellipses carry information about the Minkowski tensors;
  for more information about the  construction of the ellipses see the
  main text.   Red (medium grey)  ellipse: $V_0^{2,0}$; blue (dark  grey) ellipse:
  $V_1^{2,0}$; green (light  grey) ellipse:  $V_2^{2,0}$.  Ellipse at  the left
  hand side: $V_1^{0,2}$.
\label{fig:toy}}
\end{figure}

%%%%%%%%%%%%%%%%%%%%%%%%%%%%%%%%%%%%%%%%%%%%%%%%%%
Morphometry   deals  with   measures  for   the  content,   shape  and
connectivity of spatial patterns (``bodies'').  Consider a body $P$ in
$2$--dimensional space such  as constituted by the pixels  of a neuron
image   (see   Figures~\ref{fig:toy}   and   \ref{fig:smoothing}   for
examples).   A straightforward way  to measure  its ``content''  is to
calculate its area $V_0(P)$ or -- equivalently -- to count its pixels.
The  area   clearly  meets  the  requirement   of  motion  invariance.
Furthermore, it is additive; that is, the area of the set union $P$ of
two  bodies $P_1$,  $P_2$  can  be decomposed as
$V_0(P) = V_0(P_1)+ V_0(P_2) -  V_0(P_1\cup P_2)$.  Thus, the area can
always be calculated by summing up over local contributions from basic
building blocks  (pixels, e.g.).
% more generally, one  has to integrate
% over the parts of the body:  $V_0 = \int_P \d^2 A$.  
Finally, the area
of a  convex body can be  continuously approximated by the  areas of a
sequence of convex polygons (conditional continuity of $V_0$).
\\
There are other geometric descriptors that share these properties with
the  area. The perimeter is a case in point. However, the class
of motion-invariant, additive and conditionally continuous descriptors
is not  unbounded. Let us  point this out  in full generality  for $d$
dimensions. Consider  an arbitrary pattern $P$ that  can be decomposed
into  a  set  union  of  finitely many  convex  bodies.  According  to
{\em Hadwiger's                                             characterization
theorem}~\cite{hadwiger:1957,weil:stereology}  there  are only  $(d+1)$
linearly  independent measures $V_0(P)$,  ..., $V_{d+1}(P)$  that obey
motion-invariance,  additivity and  conditional  continuity. They  are
called  {\em (scalar)  Minkowski functionals}.  Thus, in  our  case of
$d=2$,   the  area  $V_0$,   the  perimeter   $4V_1$  and   the  Euler
characteristic  $V_2$ constitute  a  {\em complete}  family of  scalar
morphological  measures.  Note,  that  the Euler  characteristic is  a
topological invariant  and equals  the number of  connected components
minus the number of holes for patterns in $\R^2$.  The 
Minkowski  functionals were applied to  neuronal cell  classification in
\cite{Barbosa:2003a,Barbosa:2003b}.
\\
Like   the  area   $V_0$,  the   perimeter  $V_1$   and   the  Euler
characteristic $V_2$ can be  decomposed into local contributions. This
time they arise  from the boundary $\partial P$ of the body  $P$. For
smooth boundaries one has
\begin{equation}
 V_1 = \frac{1}{4} \int_{\partial P} \d^1 S,
\quad V_2 = \frac{1}{2\pi} \int_{\partial P} c\,\d^1 S , 
\end{equation}
where $c$  denotes the  curvature of $\partial  P$ and varies  as one
moves along $\partial  P$.  The factor $\frac{1}{4}$ is  a pure matter
of  convention. For pixel  sets, which  do not  have a  smooth boundary,
$V_1$ and $V_2$ can be calculated by summing up contributions from the
bonds that confine the pixels, and the corners, see~\cite{Raedt:2001}.
\\
A  natural way  of generalizing the concept of  the Minkowski
functionals is to replace the requirement of 
motion {\em in}vari\-ance by  motion {\em co}variance. Motion covariance
means that  the Minkowski valuations obey  simple transformation rules,
when the body  is moved in space: they transform  exactly as vectors or
tensors do under transformations of a coordinate system.
\\
The class of motion-covariant, additive and conditionally continuous descriptors can
be completely  characterized by a generalization  of Hadwiger's
theorem~\cite{alesker:tensor,alesker:rotation}. It turns out that they
can be reconstructed as moments of the Minkowski functionals.
\\
In two  dimensions there  are three {\em  first}-order moments  of the
Minkowski  functionals, the  so-called {\em  Minkowski  vectors}.  For
bodies with a smooth boundary, they can be represented as follows:
\begin{gather}
\V_0 = \int_P \x\,\d^2 A ,\quad \V_1 = \frac{1}{4} \int_{\partial P}
\x\ \d^1 S\notag\\
\quad \V_2 = \frac{1}{2\pi} \int_{\partial P} c \, \x \,\d^1 S , 
\end{gather}
where $\x$ denotes the position vector of the area (perimeter) element
$\d^2 A$ ($\d^1 S$) to  be integrated over. Minkowski vectors can also
be defined for pixelized images, which lack a smooth boundary.
\\
For the purposes  of our analysis, it will be  useful to normalize the
Minkowski vectors and to consider the {\em centroids}:
\begin{equation}
\p_i = \V_i/V_i\quad (i=0,1,2 \quad {\rm if}\;\; V_i\neq 0).
\end{equation}
 The  centroids specify  where some  aspect of  the  geometry (area,
 perimeter,  curvature)  is  concentrated.   Note, that  the  
 centroids $\p_i$ may,  but need not coincide with  each other. It can
 be  shown  that  all  centroids coincide  for  spherically  symmetric
 bodies.
\\
Moving  to  {\em second}-order  moments  yields  the {\em  second-rank
Minkowski tensors}.  They are  built upon the symmetric tensor product
denoted by $\x\otimes \x =: \x \x =: \x^2$. In two dimensions there  are more than three
tensors,  because,  for  $\partial P$-integrals,  instead  of  calculating
moments  with respect  to  the  spatial position  $\x$,  one may  also
consider the local normal $\n$ of the boundary, which points outwards  and is normalized to
one.\footnote{First-order moments regarding  the normal vectors always
vanish, as is shown in \cite{hadwiger:vect2}} Thus, for the
integrals $\int_{\partial P}
 \d^1 S$ and $\int_{\partial P} c
 \d^1 S$  three  types of second-order  weights for
building  moments  are  available,  viz.  $\x^r\n^s$,  where  $(r,s)=$
$(2,0)$, $(1,1)$  and $(0,2)$ (since we only  consider symmetric moments,
$\n  \x$ and $\x\n$  are identical).   Thus, altogether  the following
seven tensors can be formed:
\begin{alignat}{1}
V_0^{2,0}&= \int_P \x \x\,\d^2 A  ,\\ 
V_1^{r,s} &= \frac{1}{4} \int_{\partial P} \x^r\n^s \,\d^1 S,
\\
V_2^{r,s} &= \frac{1}{2\pi} \int_{\partial P} c \,\x^r \n^s \,\d^1 S.
\end{alignat}
In  practice, however,  we need  not consider  all of  these tensors,
because some  of them  are  linearly dependent~\cite{Schneider:2000}. It  can be
shown that only the following tensors carry independent information:
\begin{gather}
 V_0^{2,0}= \int_K \x \x\,\d^2 A  ,\\ 
V_1^{2,0} = \frac{1}{4} \int_{\partial K} \x\x \,\d^1 S,
\quad 
V_1^{0,2} = \frac{1}{4} \int_{\partial K} \n\n\,\d^1 S,\\
\quad 
V_2^{2,0} = \frac{1}{4} \int_{\partial K} c \,\x \x \,\d^1 S.
\end{gather}
In the following we will  concentrate on these tensors. They are
listed together with their names in Table~\ref{tab:tens}. The numerics
for calculating  the Minkowski valuations  for pixelized data  sets is
described in \cite{beisbart:tensor}.
\begin{table}
\begin{tabular}{|l|l|l|}\hline
Symbol & Formula & Name\\\hline\hline
$V_0$  & $\int_P \d^2 A$ &  area \\\hline
$\p_0$ & $\int_P \x \d^2 A /V_0$ & center of mass \\\hline
$V_0^{2,0}$ & $\int_P\x\x \d^2 A $ &  mass tensor \\\hline
$V_1$  & $\int_{\partial P} \d^1 S$ & length of perimeter \\\hline
$\p_1$ & $\int_{\partial P}\x \d^1 S /V_1$ & center of perimeter \\\hline
$V_1^{2,0}$ & $\int_{\partial P}\x\x \d^1 S $ &  perimeter tensor \\\hline
$V_1^{0,2}$ & $\int_{\partial P} \n\n\d^1 S $ &  $\n$-weighted perimeter tensor\\\hline
$V_2$  & $\int_{\partial P}c \d^1 S$ & Euler characteristic \\\hline
$\p_2$ & $\int_{\partial P}c \x\d^1 S /V_2$ & curvature centroid \\\hline
$V_2^{2,0}$ & $\int_{\partial P} c  \x\x \d^1 S $ &  curvature tensor \\\hline
\end{tabular}
\caption{ The Minkowski valuations used in this paper. \label{tab:tens}}
\end{table}
\\
Because of motion covariance,  the numerical values of the second-rank
Minkowski tensors depend  on the choice of the  coordinate system. But
in many applications, there is a  natural choice for the origin of the
coordinate  system. For  our neuronal  cells we  will simply  take the
position of  the soma as  the origin (in  other cases it might be  useful to
calculate the  second-rank Minkowski tensors  $V_i^{r,s}$ with respect
to the corresponding centroid $\p_i$ for $i=0,1,2$).
\\
In order to illustrate very briefly how the Minkowski valuations work for pixelized
data  sets, let  us  consider  three simple  toy  examples (some  more
examples can  be found in  \cite{beisbart:tensor}). They are  shown in
Figure~\ref{fig:toy}.   The  red (medium grey) filled square, the blue
(dark grey)
 open square and the green (light grey) x  denote  the
  centroids  $\p_0$, $\p_1$  and  $\p_2$, respectively.   The
tensors are  calculated around the center  of the black  square in the
middle of the  pixel sets as origin. The red (medium grey), blue (dark grey) and
green (light grey) ellipses
within the neurons visualize  the tensors $V_0^{2,0}$, $V_1^{2,0}$ and
$V_2^{2,0}$, respectively.   The ellipse for  the tensor $V_1^{0,2}$
is shown at the left-hand side. The equation defining
the ellipses  is always: $\x = \c +  a\left(\frac{\tau_>}{\tau_<} \cos (\phi)
\e_>+\sin(\phi)\e_<\right)$,  where  $\phi$  runs  from 0  to  $2\pi$,
$\e_{>}$ ($\e_{<}$)  is the eigenvector corresponding  to the larger
(smaller) eigenvalue $\tau_>$ ($\tau_<$) of the tensor and $\c$ is the
center  of the  soma (except  for $V_1^{0,2}$; its
ellipse is shifted  to the edge of the panels).   So the axis ratios
of the  ellipses are the ratios  of the eigenvalues,  and the ellipses
point  into   the  direction  of  the  eigenvector   with  the  larger
eigenvalue.   The  size  of  the  ellipses  does  not  carry  specific
information because of the free scale factor $a>0$.
\\
 In the  top panel of Figure~\ref{fig:toy} the pixel  set displays an axial  symmetry and is
almost point  symmetric. Accordingly, the centroids are  very close to
each  other;  they  fan  out  along the  symmetry  axis.  The  tensors
$V_i^{2,0}$  align perpendicular  to  the symmetry  axis, because  the
whole pixel set  is more elongated along the  horizontal axis. The
tensor ellipses for the mass tensor $V_0^{2,0}$ and the perimeter tensor $V_1^{2,0}$ almost coincide,
whereas the ellipse corresponding to $V_2^{2,0}$ is a bit more
elongated.  The reason is that the corners, which play an important
role for the curvature tensor
$V_2^{2,0}$ are further away from the middle black square, which only
contributes to $V_0^{2,0}$ and $V_1^{2,0}$. 
\\
For the middle panel, the  figure has been slightly modified: in order
to  destroy the symmetry,  we rearranged  one of  the ``arms''.   As a
consequence, the average pixel is  lower down than in the first panel,
so all centroids move downwards.  The effect is most prominent for the
curvature  centroid  $\p_2$,  because  it  depends  on  corners, some of  which
disappear for  the rearranged  dendrite.  Note, furthermore,  that the
 centroids span a non-degenerate triangle, a fact that can be
taken as indicating asymmetry. The  lack of symmetry is also reflected
by  the  tensor  ellipses,  which  are not  parallel  any  more.   Note,
furthermore,  that  the ratios  between  the  bigger  and the  smaller
eigenvalues are larger for the second  pixel set. The reason is that --
due  to  the ``movement''  of  the upper  right  arm  -- the  vertical
extension of  the pixel  set shrinks on  average, such that  the whole
body is more elongated.
\\
The bottom  panel shows a  variation of the  body in the  middle panel,
where  two   holes  have  been   added.  This  results  in   an  Euler
characteristic of $-1$.  There is no major effect  for both $\p_0$ and
$\p_1$  and the  related tensors.  But for  $\p_2$ a  big jump  can be
observed,  and the ellipse  for the  curvature tensor  $V_2^{2,0}$ is
twisted and more  elongated.  The position of $\p_2$  can be explained
as  follows: The  hole at  the right-hand  side makes  a  big negative
contribution  to $\V_2$.   So, if  $\V_2$ is  calculated  around the
center of the black square, it points to the left hand side.  But
since the  Euler characteristic  $V_2$ itself  is negative,  $\p_2$ is
bounced back to   the
right hand side  due to its
normalization through $V_2$. For the curvature tensor ellipse there is some kind of
repulsion from the right hole, because this hole makes a big negative
contribution to the tensor; the effect of the other hole is much
smaller because it is closer to the soma. 
\\
The tensor   $V_1^{0,2}$ is shown  at the  left hand
side.  It always aligns parallel to  the grid axis,
the reason for  this being that it crucially  depends on normals that
can only point into four directions for a square lattice.\footnote{
For an  elementary proof, you can  start with a single  pixel and then
use additivity.
}  The shape  of the $V_1^{0,2}$ ellipse can be
understood as follows: The eigenvalues of $V_1^{0,2}$ count the number
of bonds  with horizontal or  vertical normals, respectively.  For all
toy  examples  there are  more  vertical  normals,  so the  tensor  is
anisotropic.   By  moving from  the  top  to  the middle  panel,  more
horizontal than vertical normals are destroyed; in this way the tensor
becomes even more anisotropic.
\\
Let us conclude  this part by adding two  comments.  First, note, that
by considering the  eigenvalues of a tensor with  respect to an origin
which is given by the body itself, motion-invariance is regained.  But
does this  mean that  we have been  returning to the  scalar Minkowski
functionals themselves?  The answer  is no.  Additivity has been lost,
because  forming eigenvalues  is not  a  linear operation,  and, as  a
consequence,  the   eigenvalues  of  a  Minkowski   tensor  cannot  be
decomposed in  the same way as  the area is. So  we have significantly
extended  the   Minkowski  framework  without  having   given  up  its
conceptual foundations.
\\
Second, there is a natural extension of
our framework to three-dimensional neuron data.
%
%%%%%%%%%%%%%%%%%%%%%%%%%%%%%%%%%%%%%%%%%%%%%%%%%%%%%%%%%%%%
\section{The analysis of pixelized neuron data}
\begin{figure}
\centering\includegraphics[width=8cm]{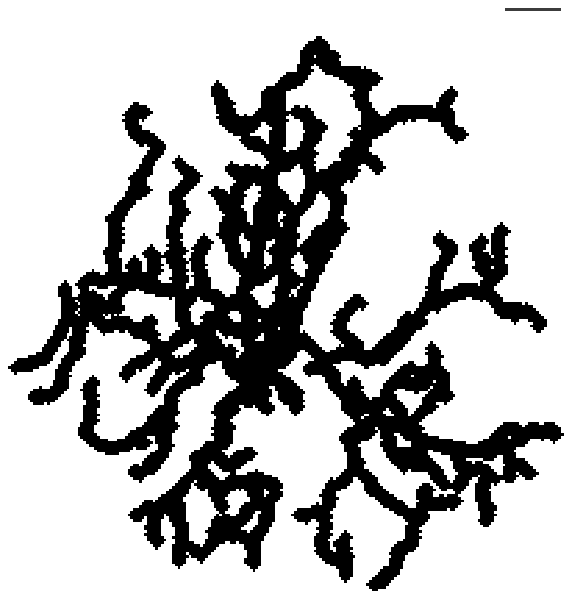}
\centering\includegraphics[width=8cm]{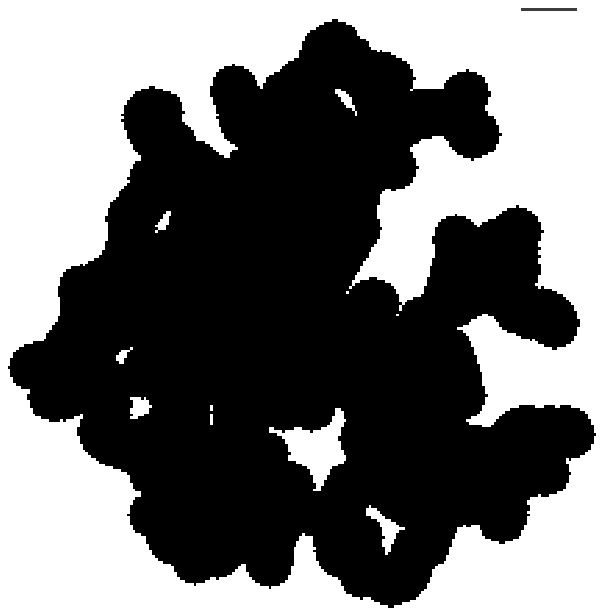}
\caption{Two  smoothed versions of  the $\lambda$-neuron.   Top panel:
  smoothing length: 2 pixels. Bottom panel: 8 pixels. The pictures are
  based  on  data obtained by~\cite{Masland:2001}  (their  figure  no.  5,
  copyright                        permission                       by
  \href{http://www.nature.com/neuro}{Nature Neuroscience}). \label{fig:smoothing} }
\end{figure}

\begin{figure}
\centering\includegraphics[width=8.4cm]{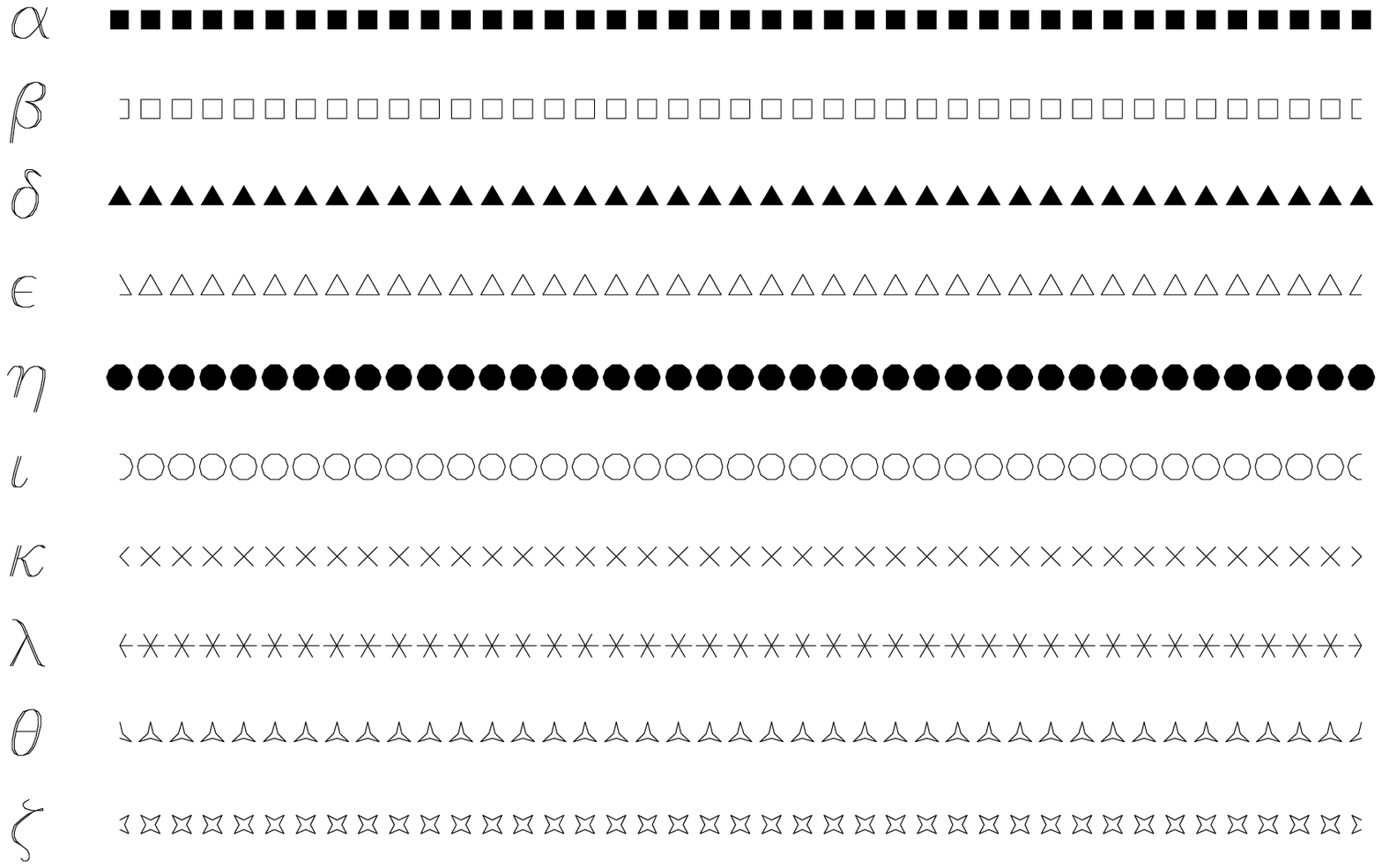}
\centering\includegraphics[width=8.4cm]{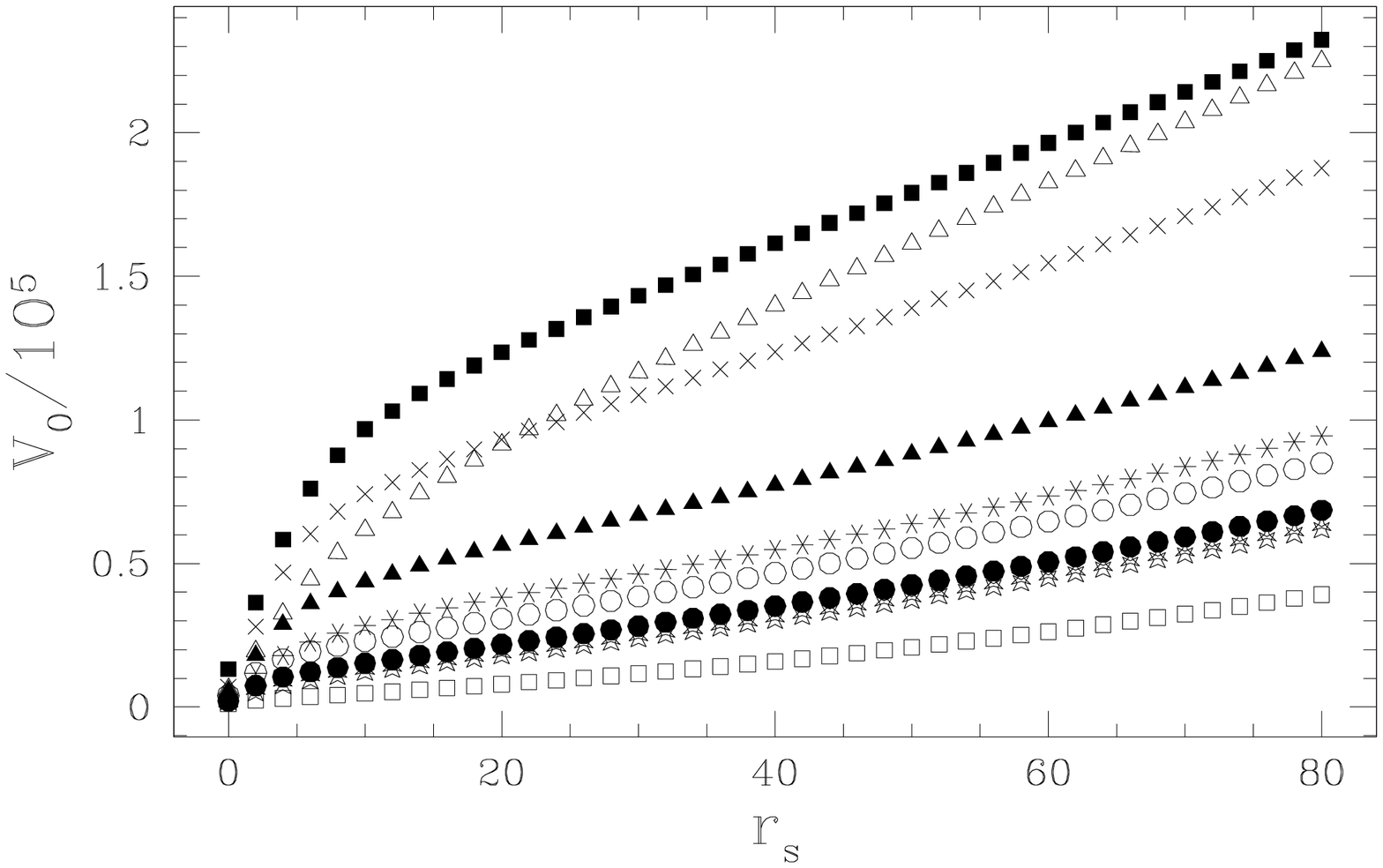}
\caption{Top  panel: the  point styles  to be  used for  the different
  types  of  neuronal cells.  Bottom  panel:  the  volume $V_0$  as  a
  function     of      $r_s$     for     a      large     range     of
  $r_s$-values. }\label{fig:smooth_sc0}
\end{figure}

\begin{figure}
\centering\includegraphics[width=8.4cm]{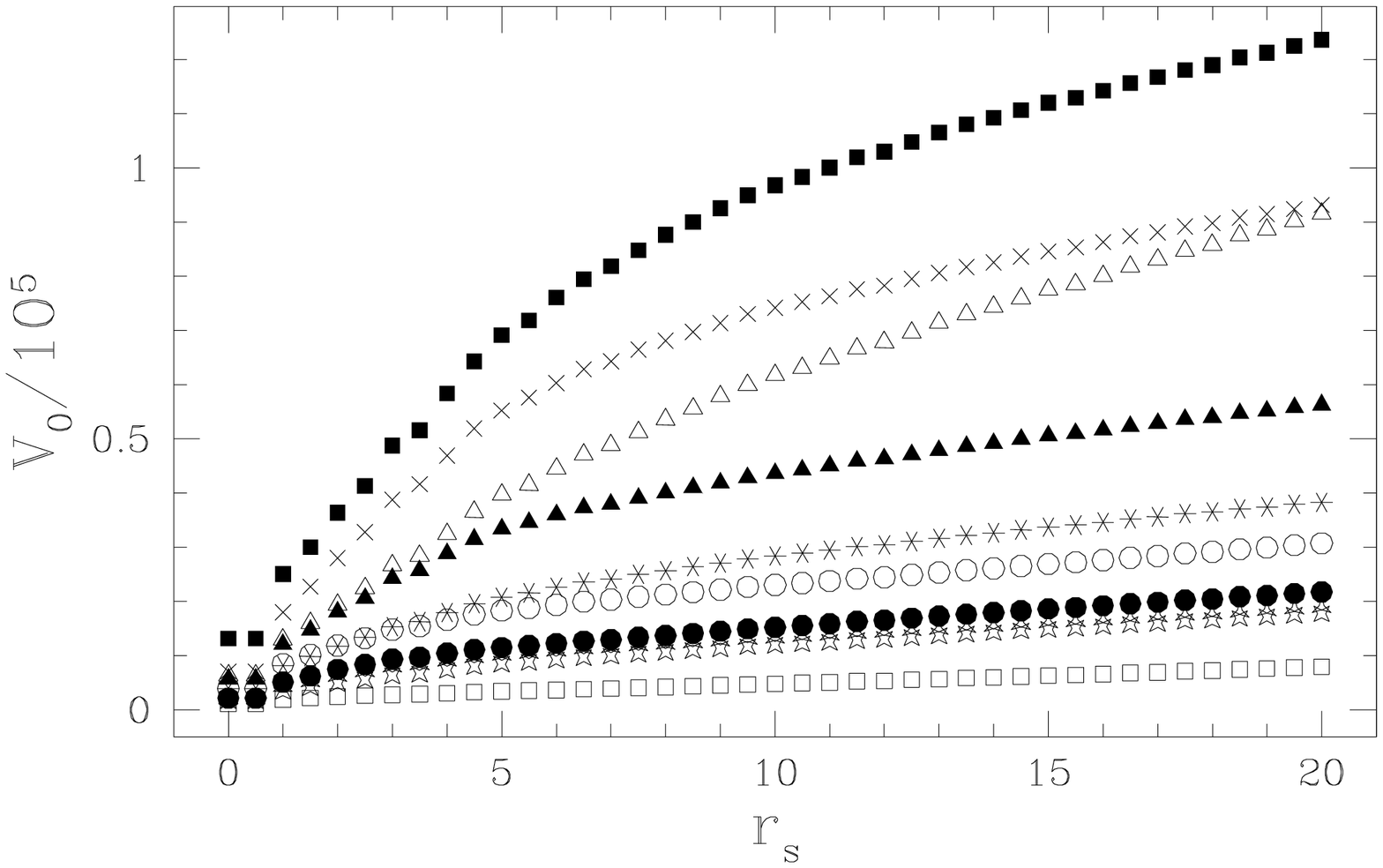}
\centering\includegraphics[width=8.4cm]{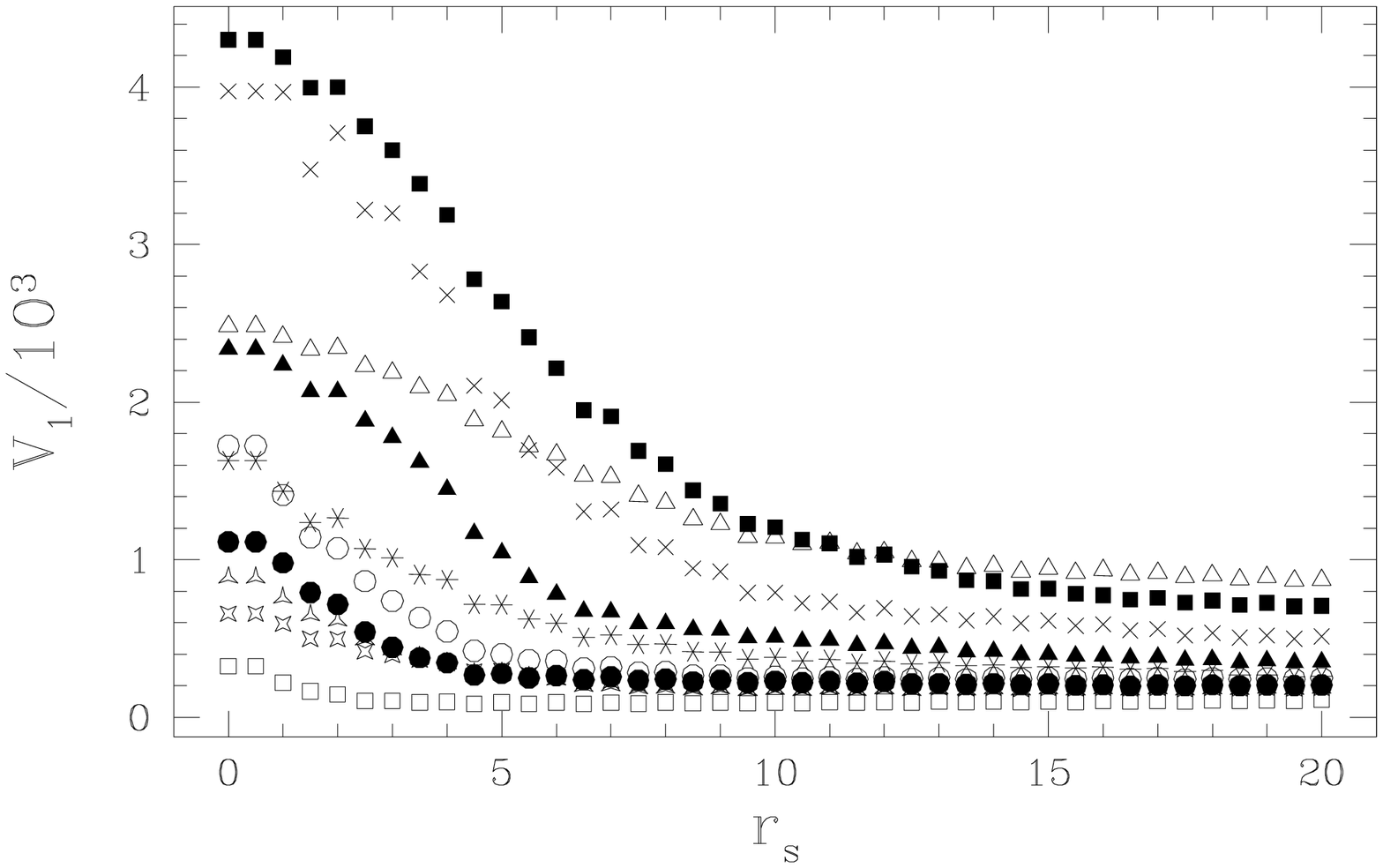}
\centering\includegraphics[width=8.4cm]{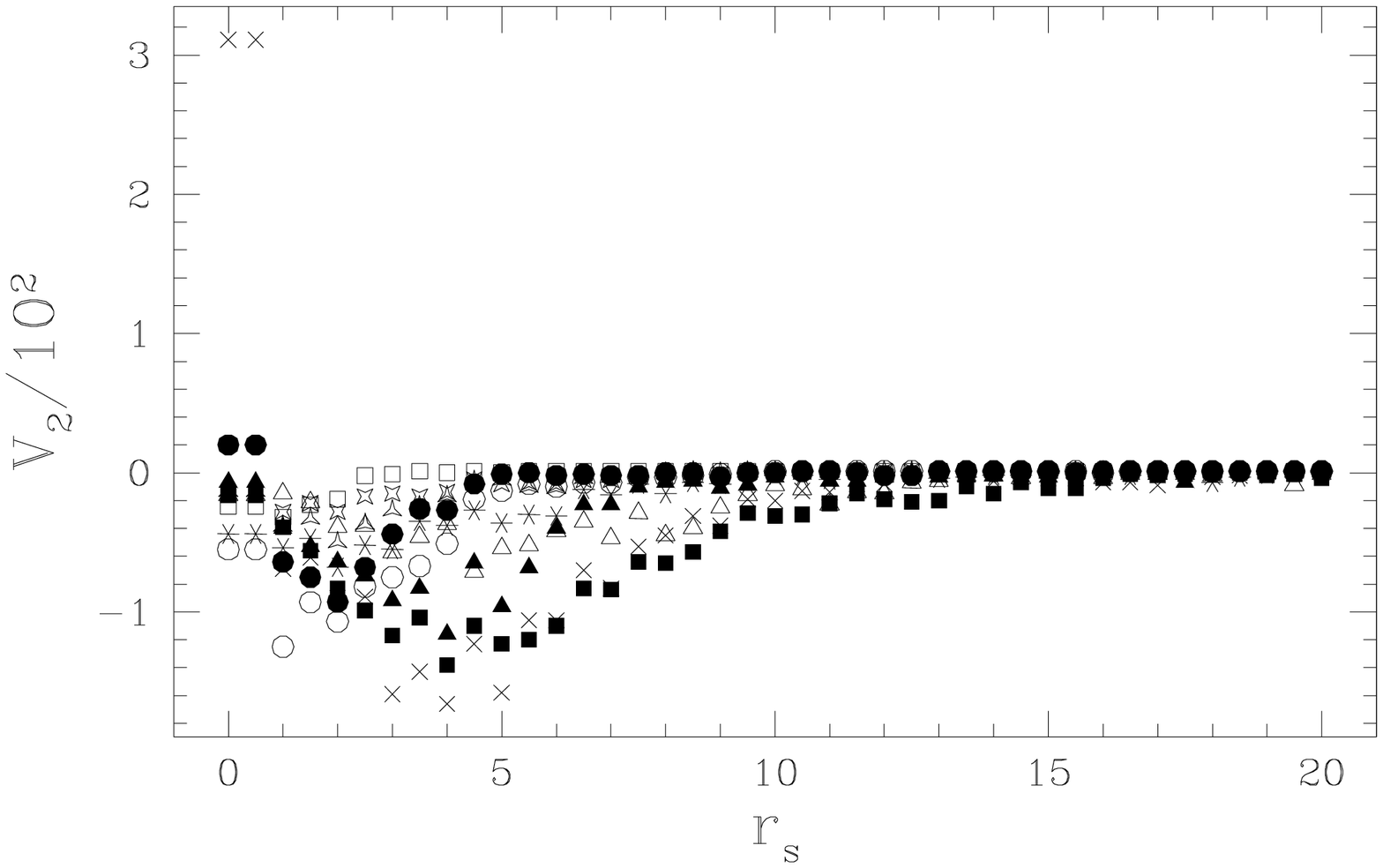}
\caption{The scalar Minkowski functionals as a function of the
  smoothing length for all cells. (Note, that in the first panel the
  curve for the $\theta$ cell is between the curves for the $\eta$ and
  $\zeta$ types.) }\label{fig:smooth_sc}
\end{figure}

\begin{figure}
\centering\includegraphics[width=8.4cm]{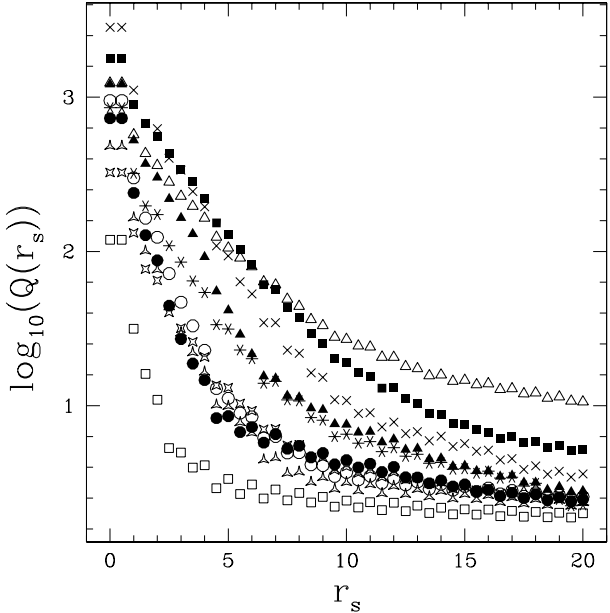}
\caption{The logarithm  of $Q$ (a variety of  the isoperimetric ratio)
as a function of $r_s$.}\label{fig:iso}
\end{figure}

\begin{figure}
\centering\includegraphics[width=8.4cm]{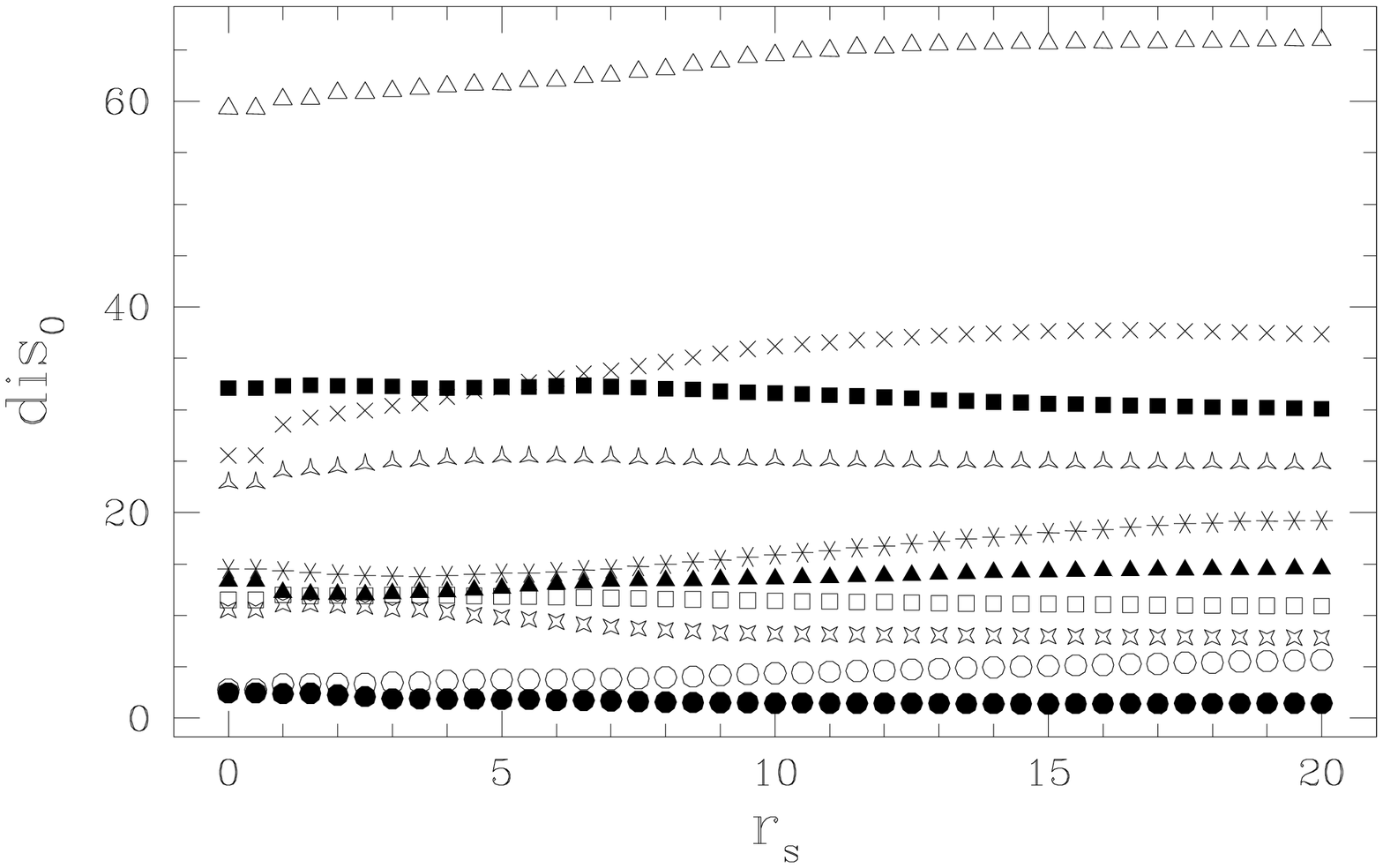}
\centering\includegraphics[width=8.4cm]{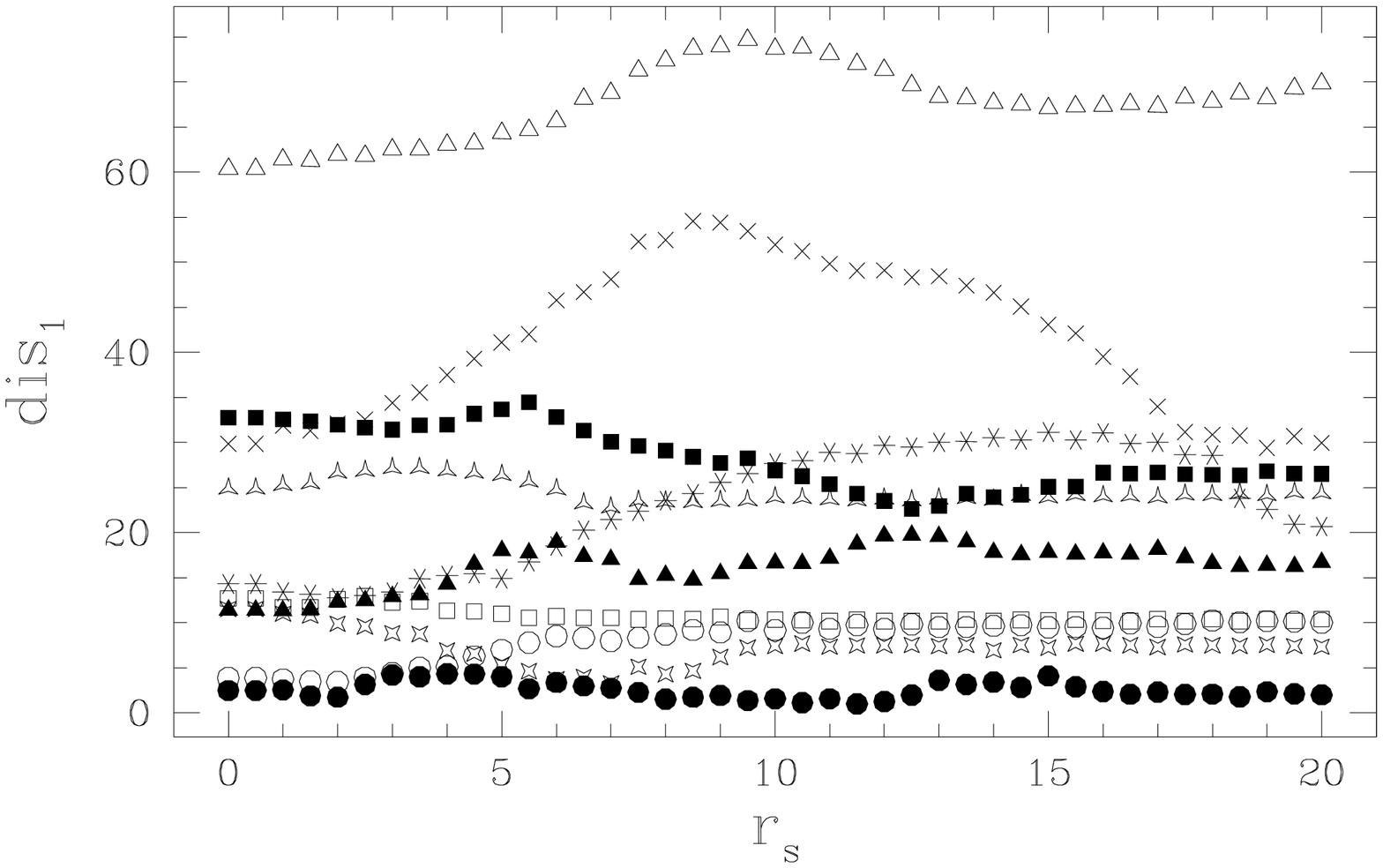}
\caption{The distance soma  -- $\p_0$ (top panel) and  soma -- $\p_1$
  (bottom  panel)   as   a   function  of   the   smoothing   length.}
  \label{fig:smooth_dis}
\end{figure}

\begin{figure}
\centering\includegraphics[width=8.4cm]{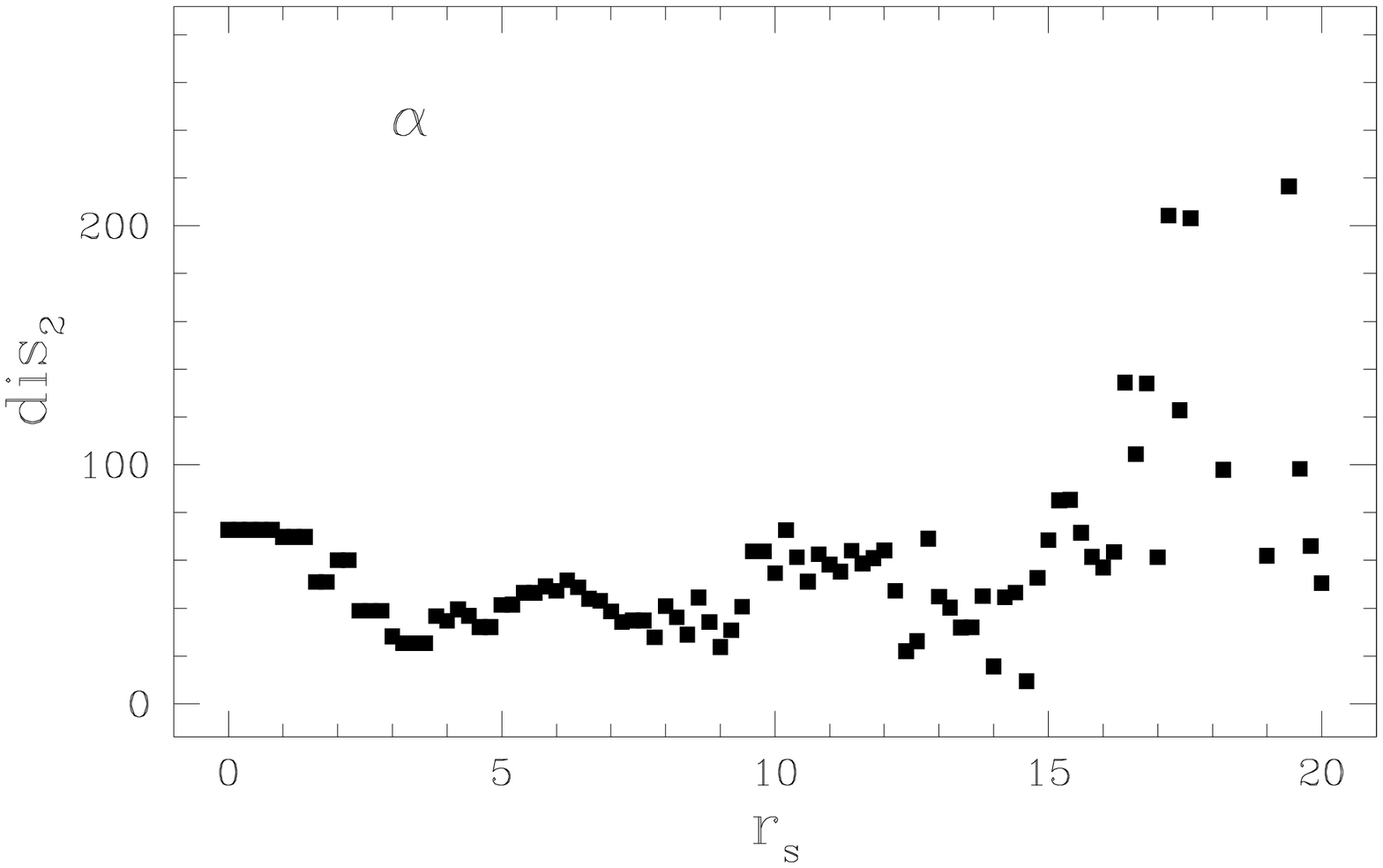}
\centering\includegraphics[width=8.4cm]{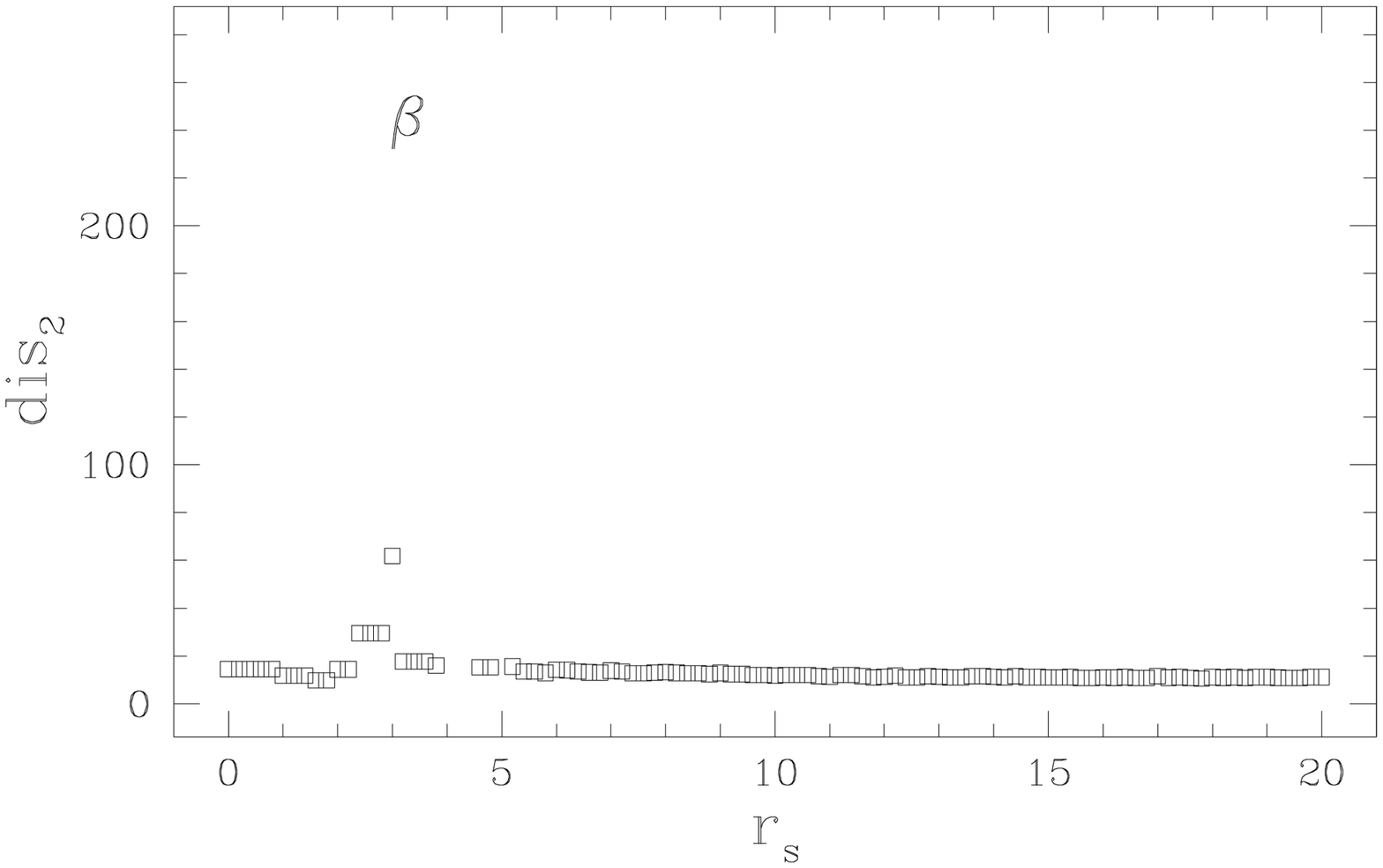}
\centering\includegraphics[width=8.4cm]{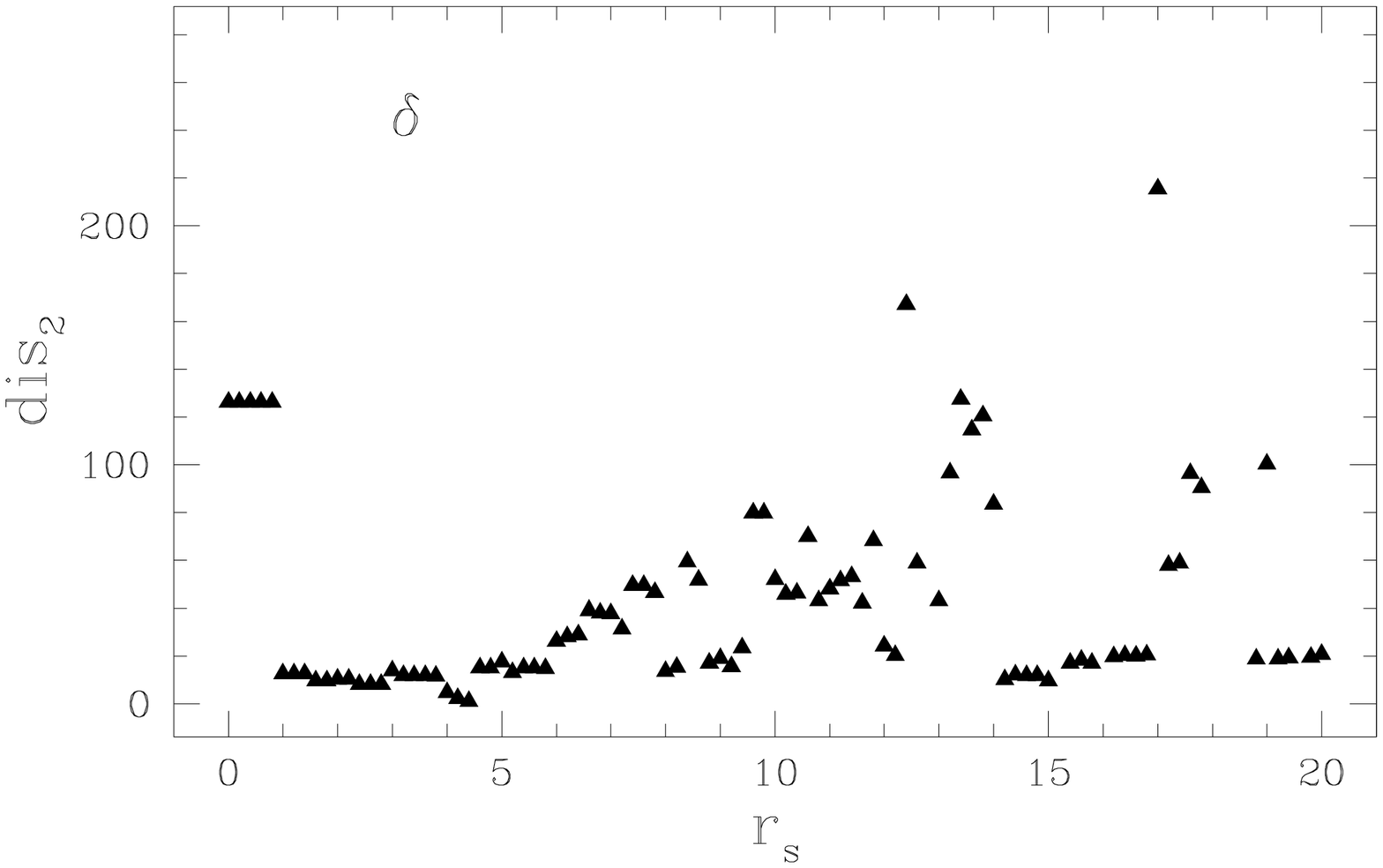}
\centering\includegraphics[width=8.4cm]{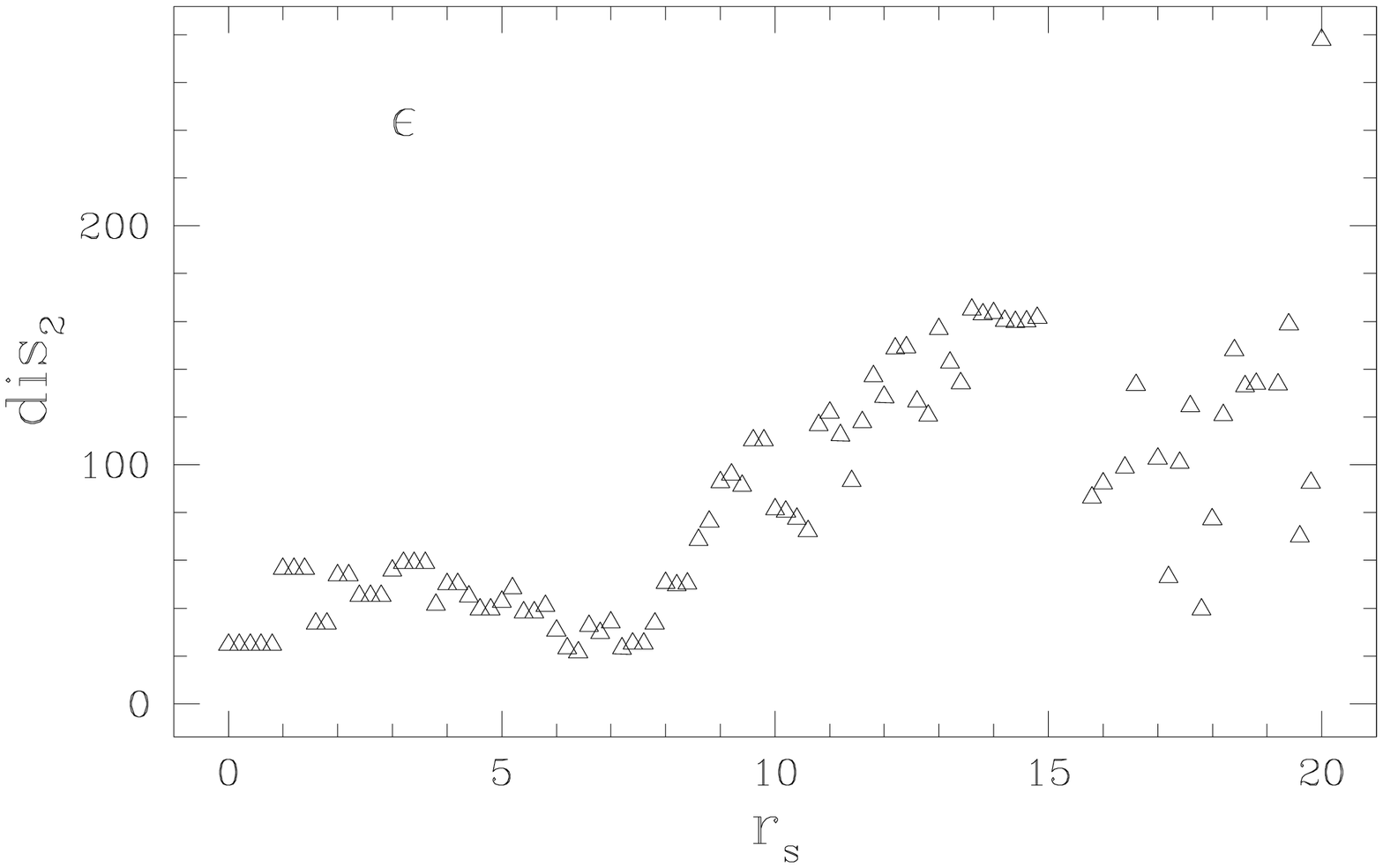}
\caption{The distances soma  -- $\p_2$ as a function  of the smoothing
  length for four  cells. If $\p_2$ is not defined  for some $r_s$
  (because of $V_2=0$), no
  point is shown.} \label{fig:smooth_dis2}
\end{figure}

\begin{figure}
\centering\includegraphics[width=8.4cm]{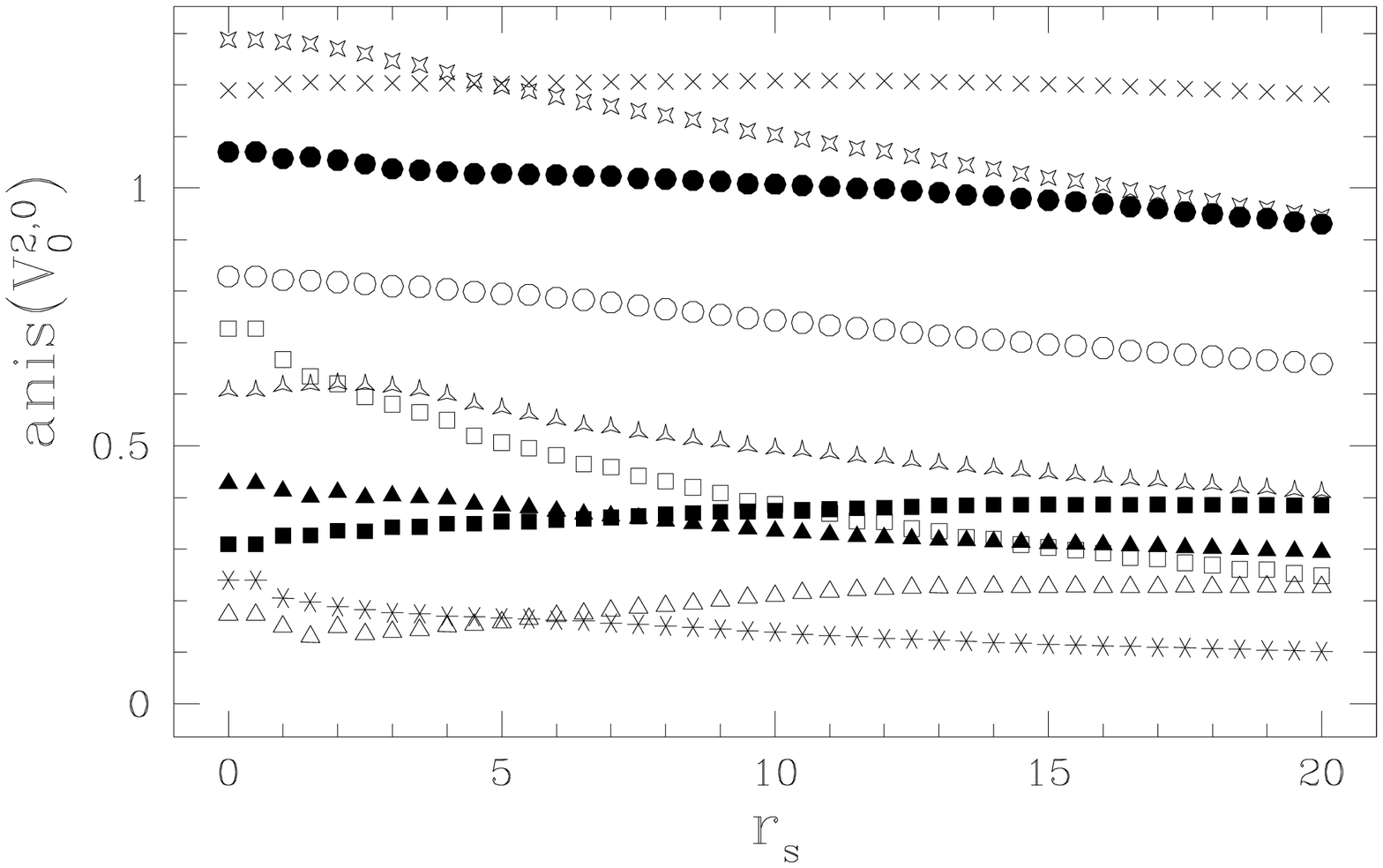}
\centering\includegraphics[width=8.4cm]{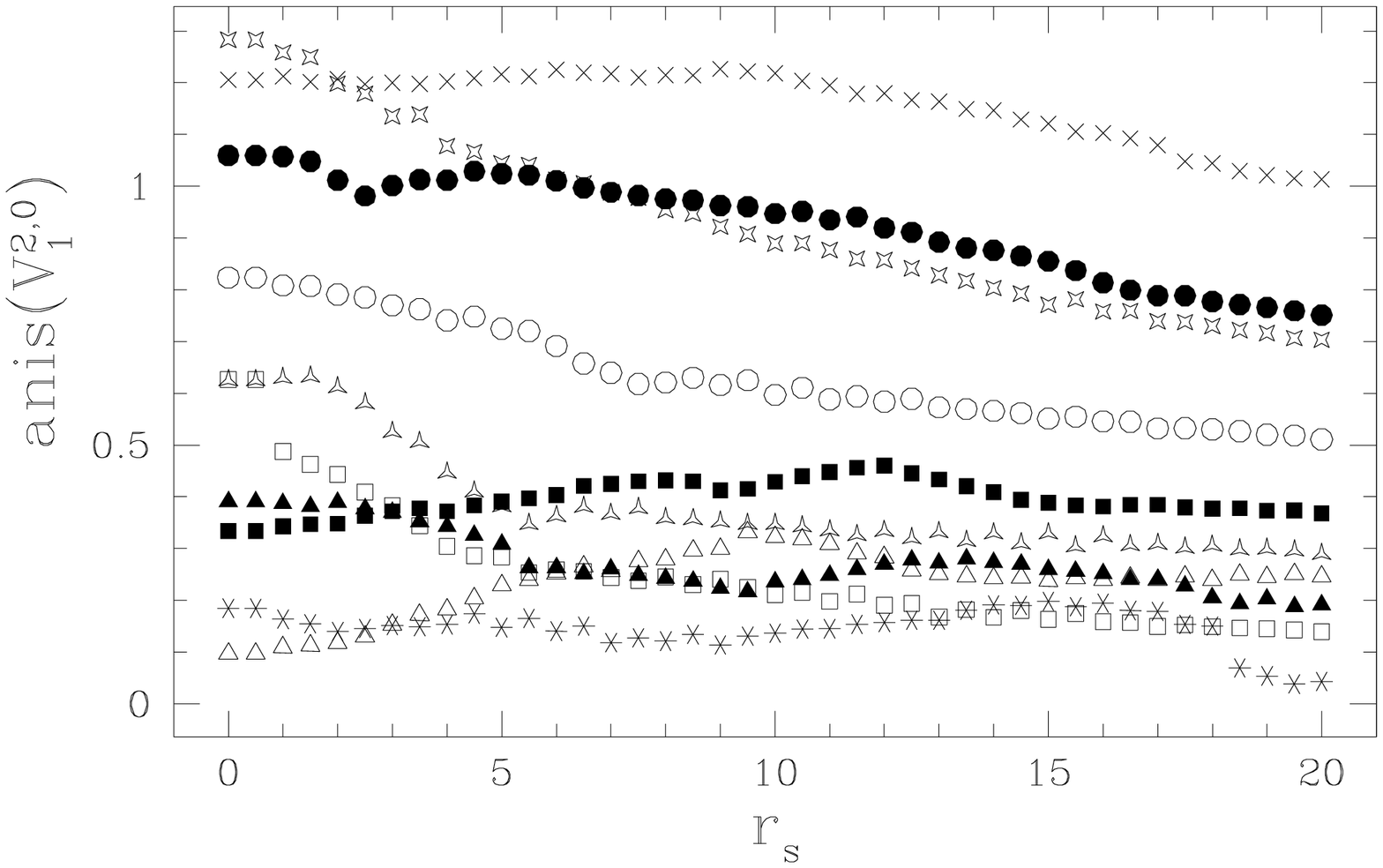}
\caption{The  anisotropy  parameters  derived  from  the  mass  tensor
  $V_0^{2,0}$ (top panel) and the perimeter tensor $V_1^{2,0}$ (bottom
  panel)      as      a       function      of      the      smoothing
  length.}\label{fig:smooth_anis}
\end{figure}

\begin{figure}
\centering\includegraphics[width=8.4cm]{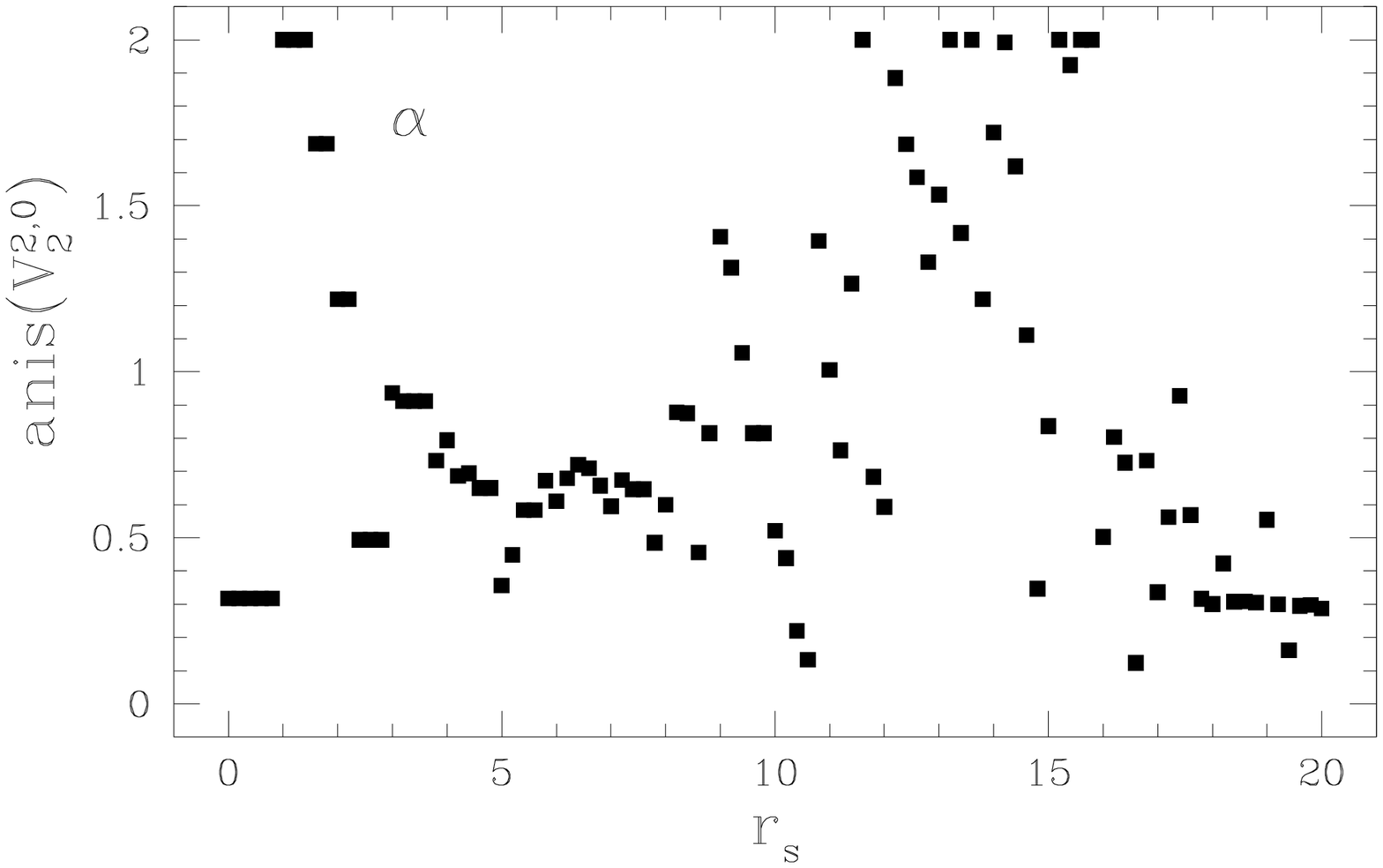}
\centering\includegraphics[width=8.4cm]{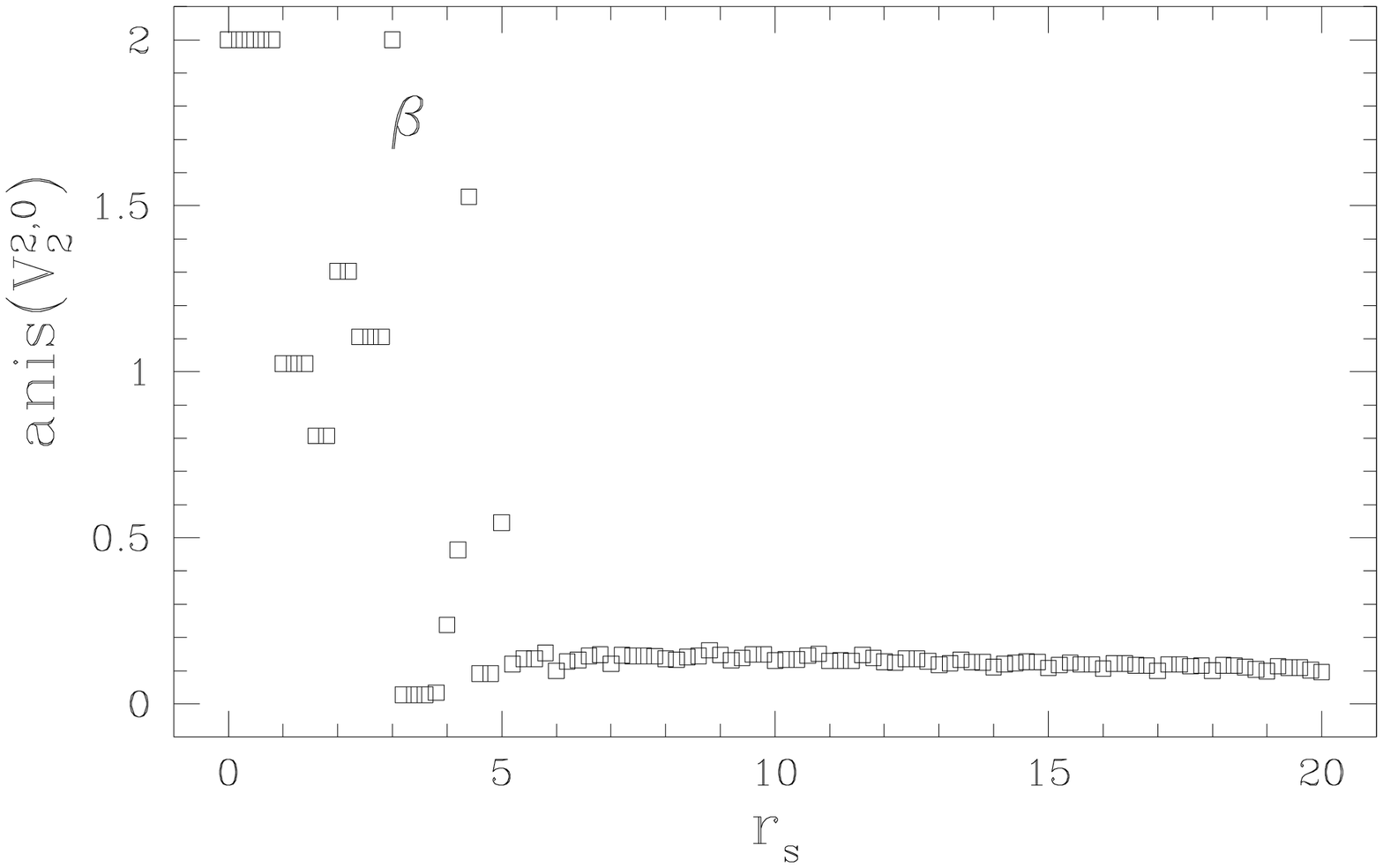}
\caption{Another anisotropy  parameter (derived from  $V_2^{2,0}$) for
  two     cells     as     a     function    of     the     smoothing
  length.}\label{fig:smooth_anis4}
\end{figure}

\begin{figure}
\centering\includegraphics[width=8.4cm]{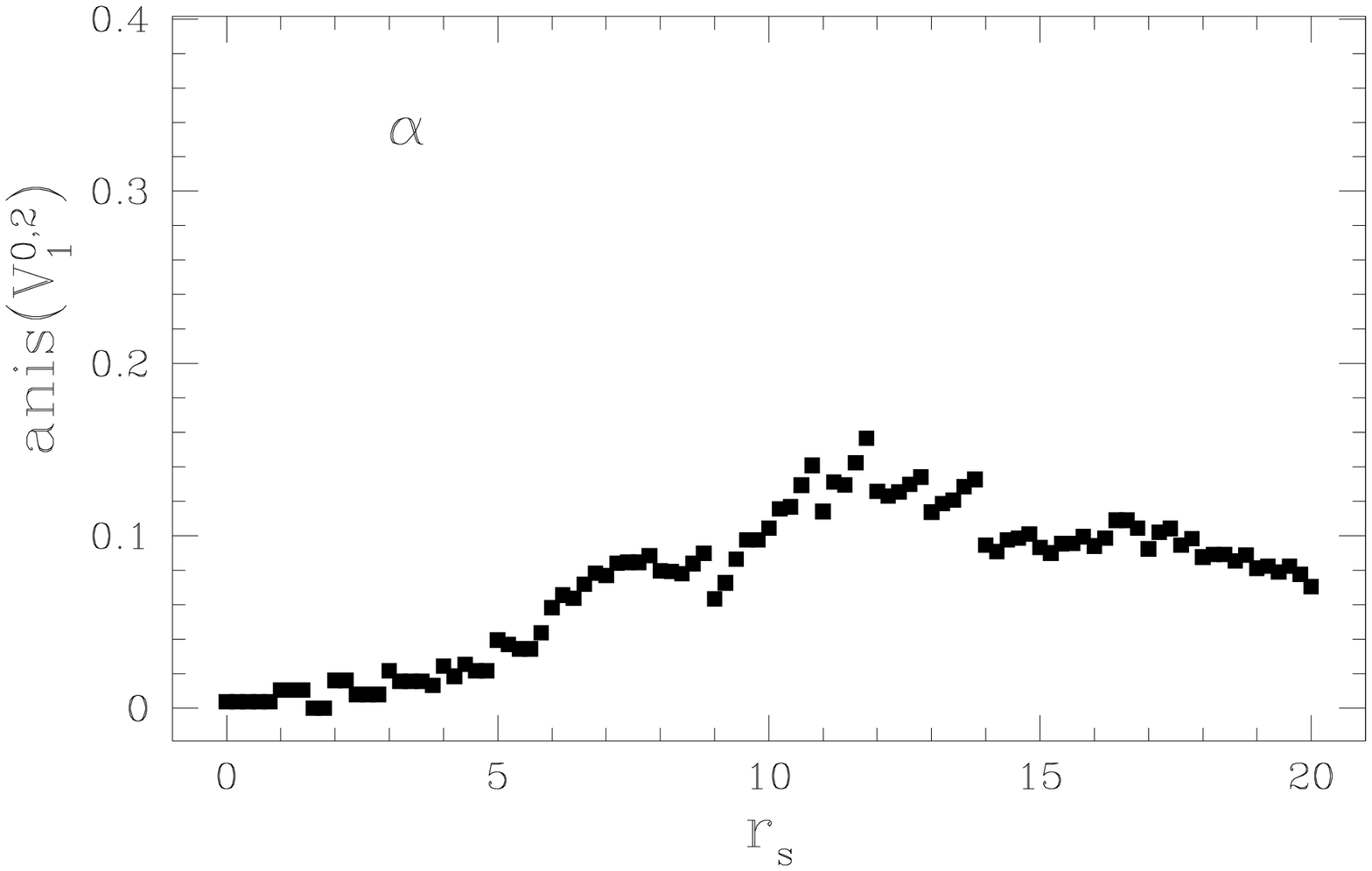}
\centering\includegraphics[width=8.4cm]{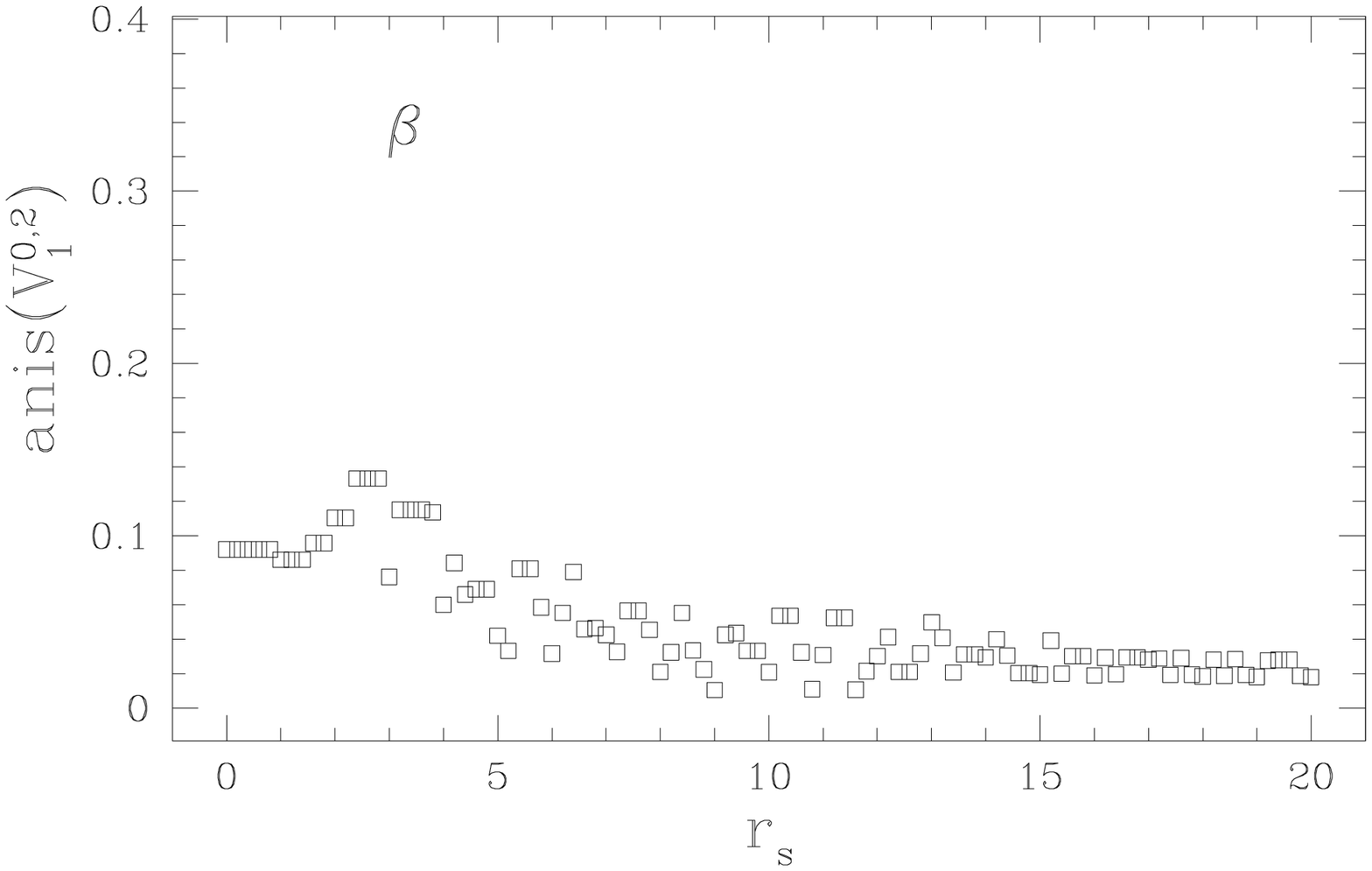}
\centering\includegraphics[width=8.4cm]{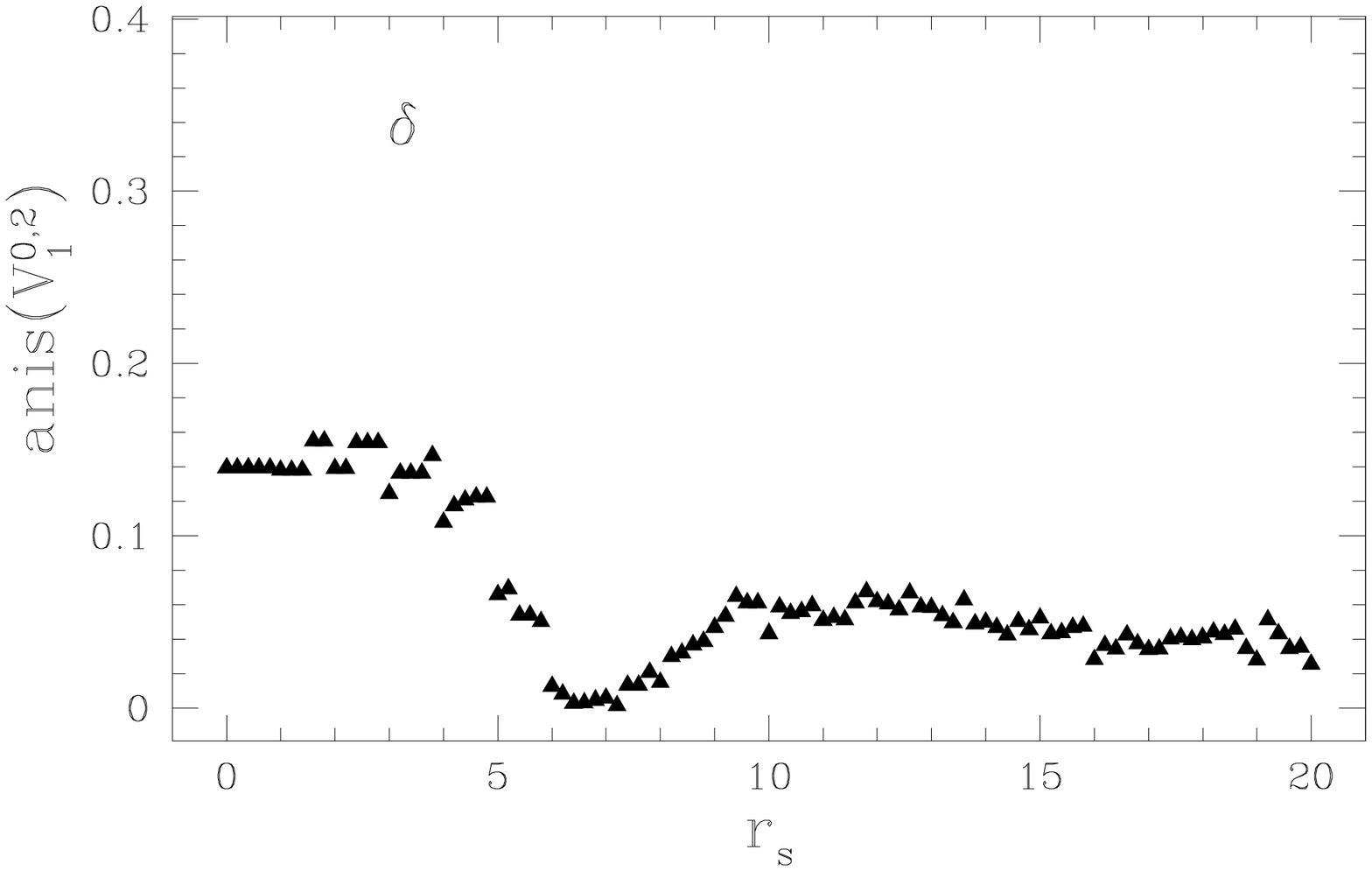}
\centering\includegraphics[width=8.4cm]{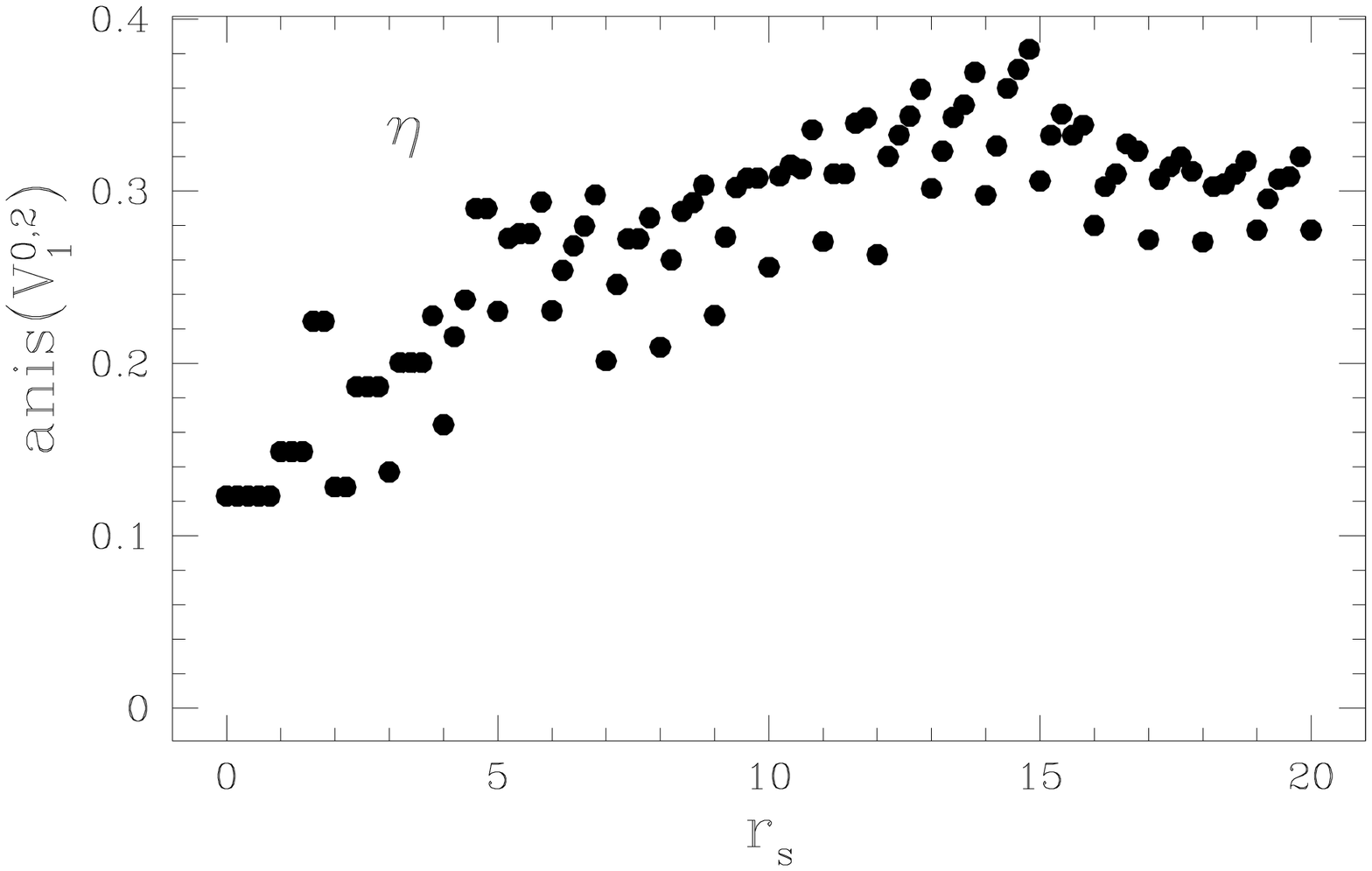}
\caption{The  anisotropy parameter derived  from $V_1^{0,2}$  for four
  cells      as       a      function      of       the      smoothing
  length.}\label{fig:smooth_anis5}
\end{figure}

\begin{figure}
\centering\includegraphics[width=8.4cm]{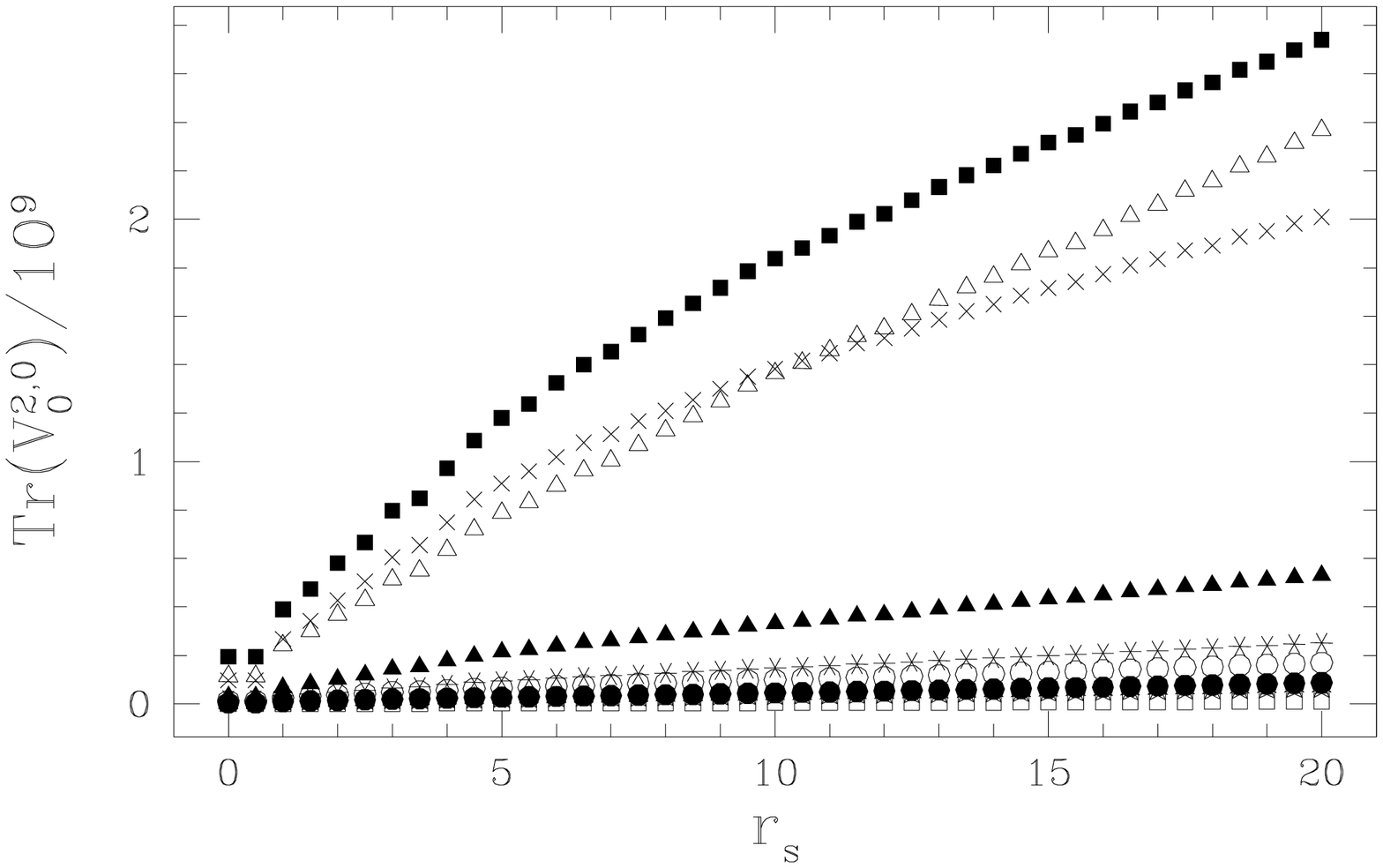}
\centering\includegraphics[width=8.4cm]{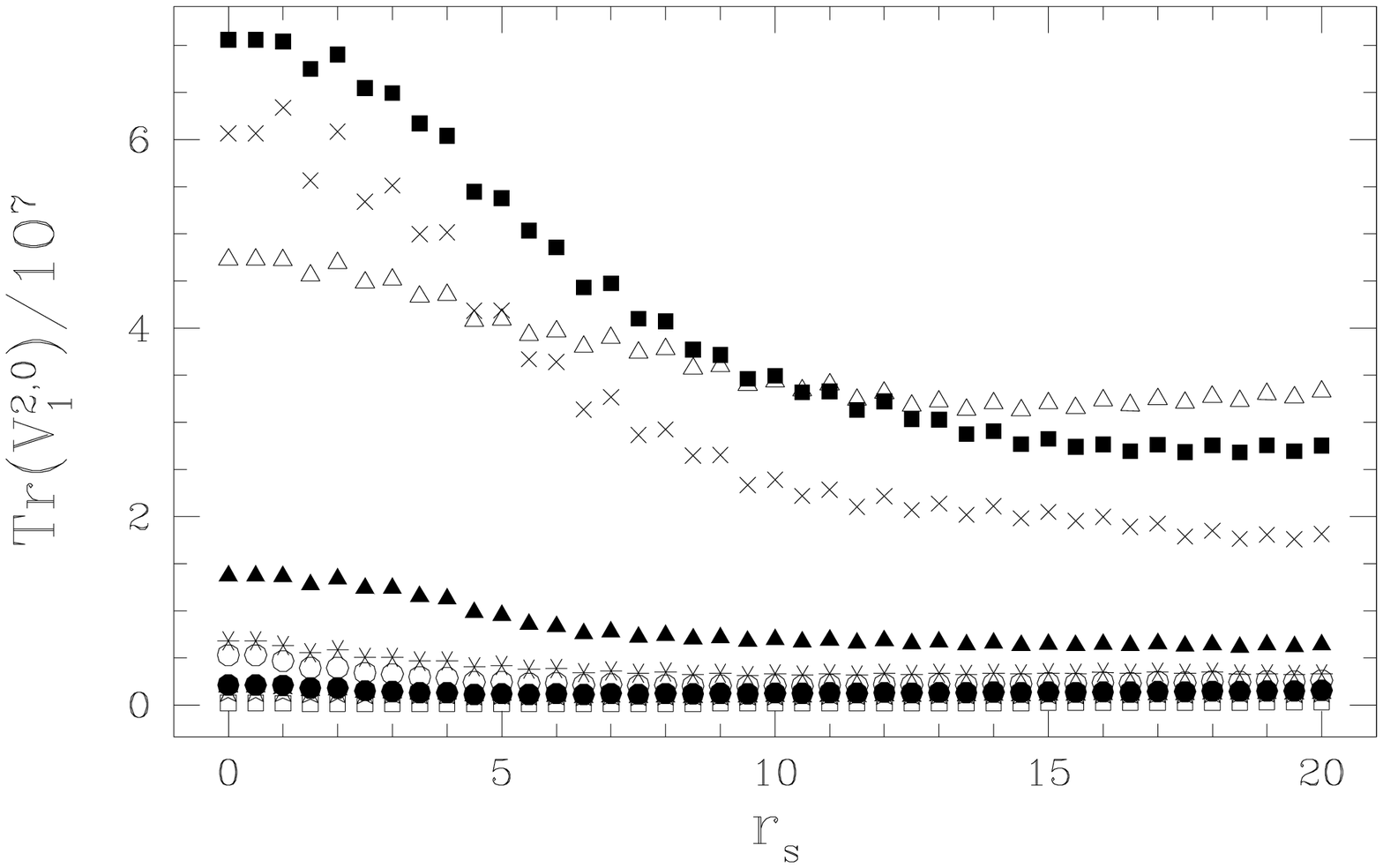}
\centering\includegraphics[width=8.4cm]{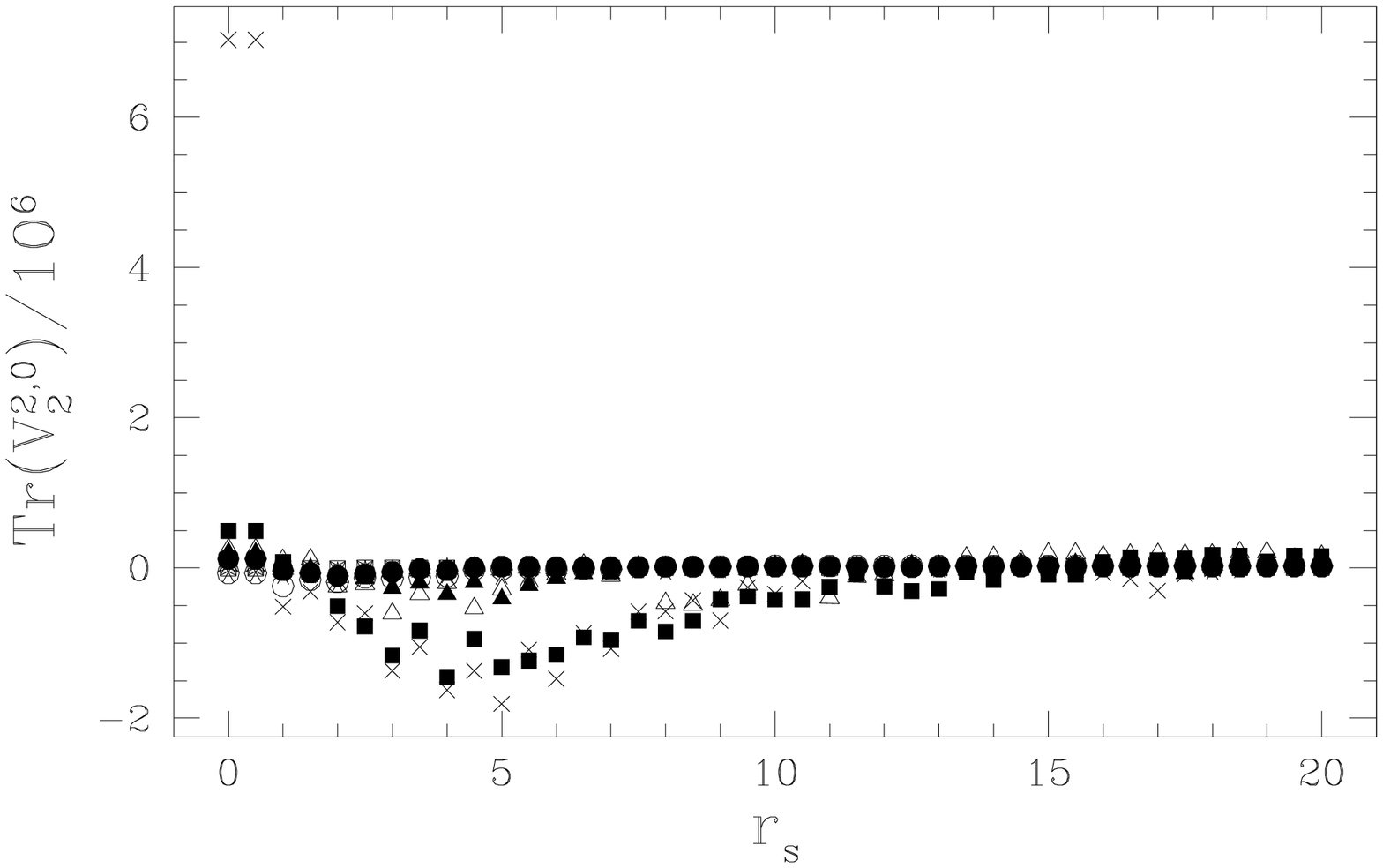}
\caption{The  traces of the  tensors $V_i^{2,0}$  for $i=0,..,2$  as a
function of the smoothing length.}\label{fig:smooth_tr}
\end{figure}

\begin{figure}
\centering\includegraphics[width=8.4cm]{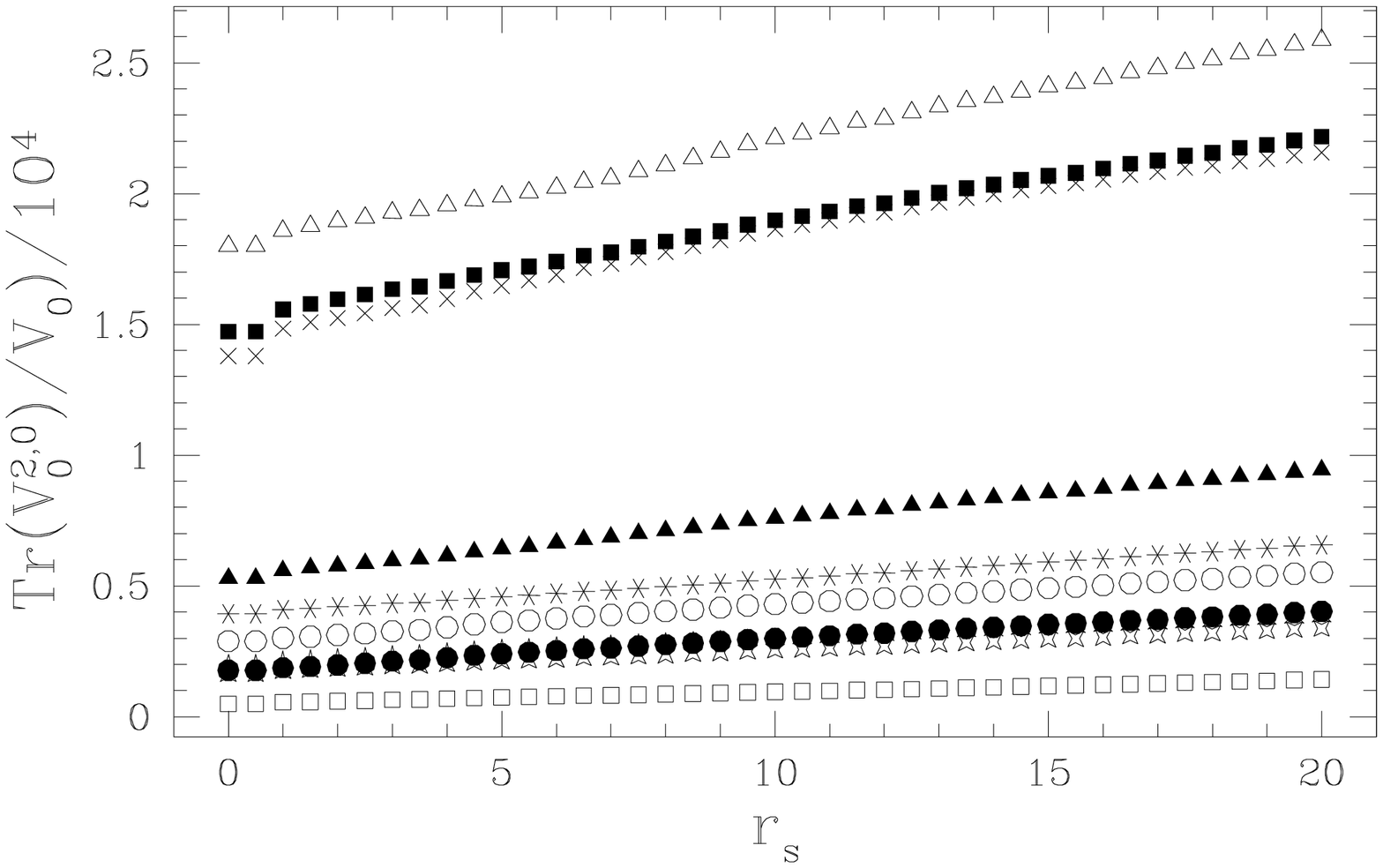}
\centering\includegraphics[width=8.4cm]{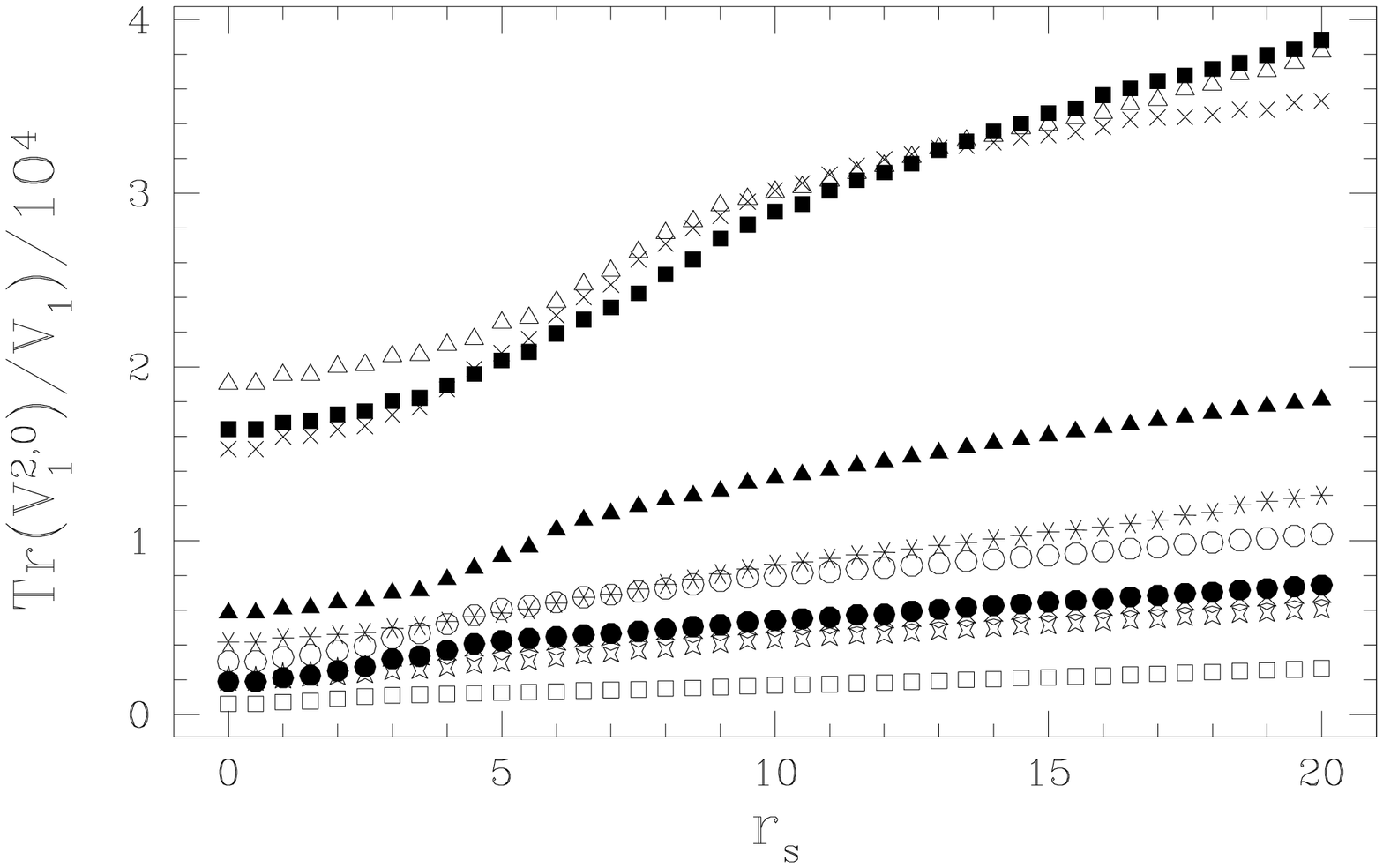}
\centering\includegraphics[width=8.4cm]{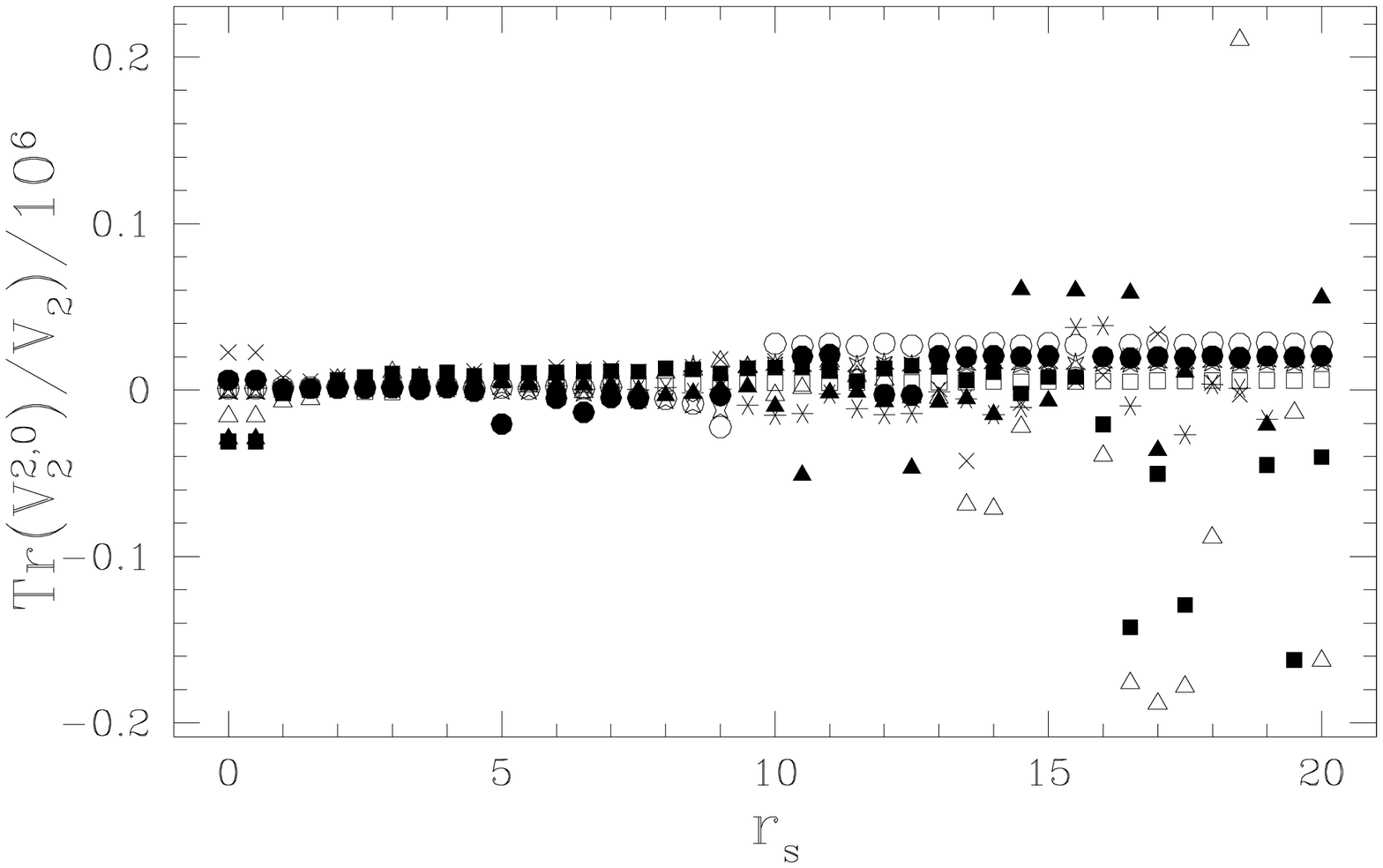}
\caption{The traces  of $V_0^{2,0}$, $V_1^{2,0}$,  and $V_2^{2,0}$, now
  normalized  by  the corresponding  scalar  $V_0$,  $V_1$, or  $V_2$,
  respectively. If $V_2=0$  for some $r_s$, no data  point is shown at
  all. }\label{fig:smooth_trn}
\end{figure}

%%%%%%%%%%%%%%%%%%%%%%%%%%%%%%%%%%%%%%%%%%%%%%%%%%%%%%%%%%%%
\paragraph{Data.}
We analyze two-dimensional neuron  data made a\-vailable by the courtesy
of Prof.   Berson~\cite{Masland:2001}. We have pixelized  maps of ten
neurons.   They  are  assigned  different  types  ($\alpha$,  $\beta$,
$\delta$, $\epsilon$, $\eta$,  $\iota$, $\kappa$, $\lambda$, $\theta$,
$\zeta$).  The neuron  maps greatly  differ  in terms  of scale.  Each
neuron can be thought of as a subset of filled pixels within a
square  lattice.  Not all of  the neuron pixel  sets are connected;
some of them  consist of disconnected parts.  This  is probably due to
an artifact  of the  neuron observations.  We  will therefore  apply a
simple smoothing.
\paragraph{Method.}
For each cell  we construct parallel sets with a  ball of radius $r_s$
on a  pixel approximation.  The parallel set  $P_{r_s}$ of a  body $P$
comprises all points $\x$ such  that the distance between $\x$ and $P$
is    $r_s$   at    most.    The    smoothing   is    illustrated   in
Figure~\ref{fig:smoothing}, where  the $\lambda$-neuron is considered.
In the sequel, the smoothing length will be varied and used as a diagnostic
parameter. It serves to probe structures at different scales.
\\
\noindent For each neuron that has been smoothed with a particular
smoothing length, we calculate the scalar Minkowski functionals, the
 centroids and the second-rank tensors.  For the tensors we
choose the center of the soma as a natural origin. The soma and its
center are identified visually, in an interactive way.
\begin{figure}
\centering\includegraphics[scale=0.55]{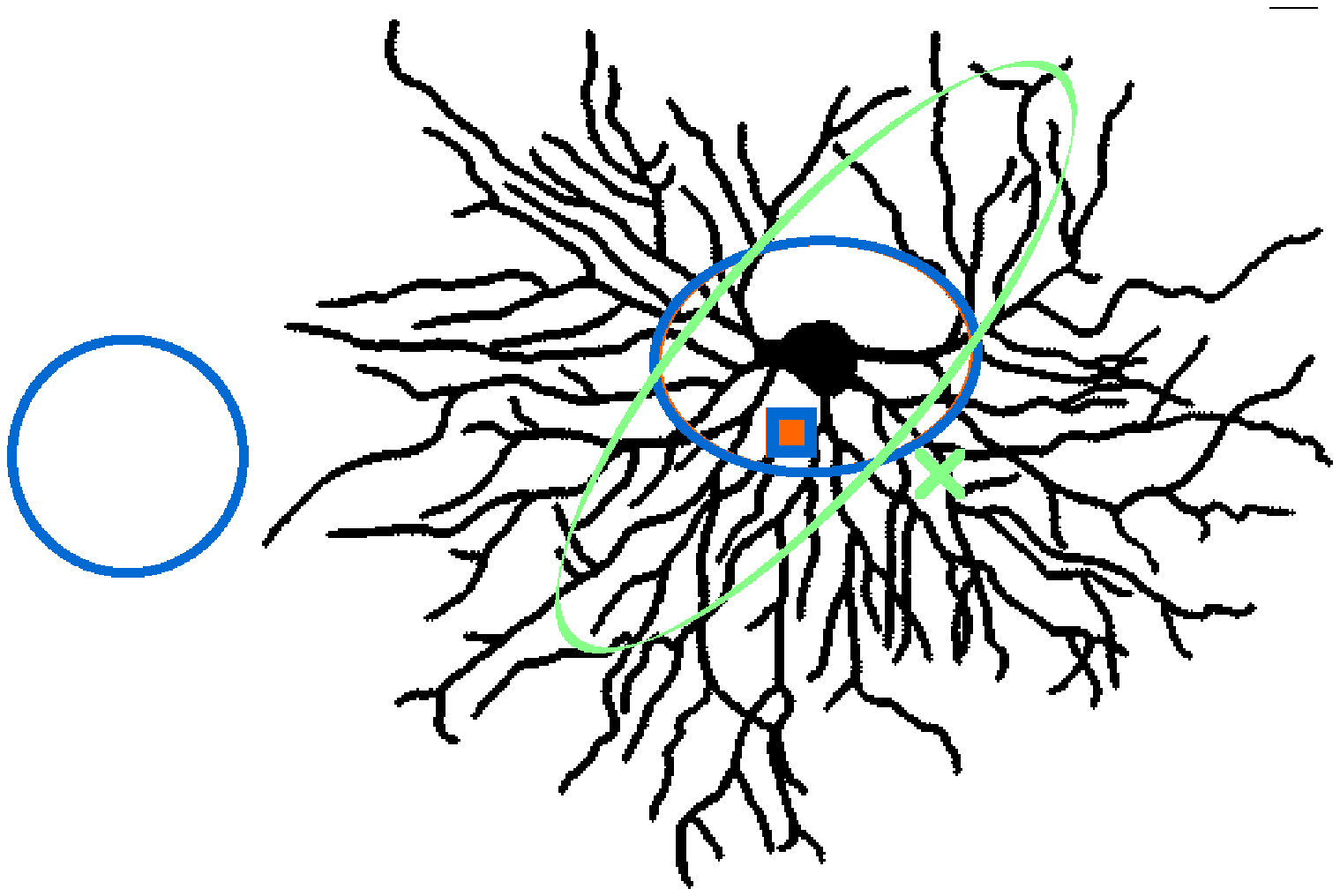}
\begin{minipage}[h]{.22\linewidth}
\centering\includegraphics[scale=0.55]{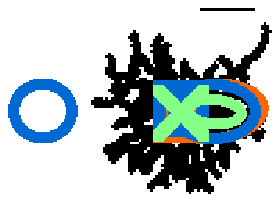}
\end{minipage}
\begin{minipage}[h]{.77\linewidth}
\centering\includegraphics[scale=0.55]{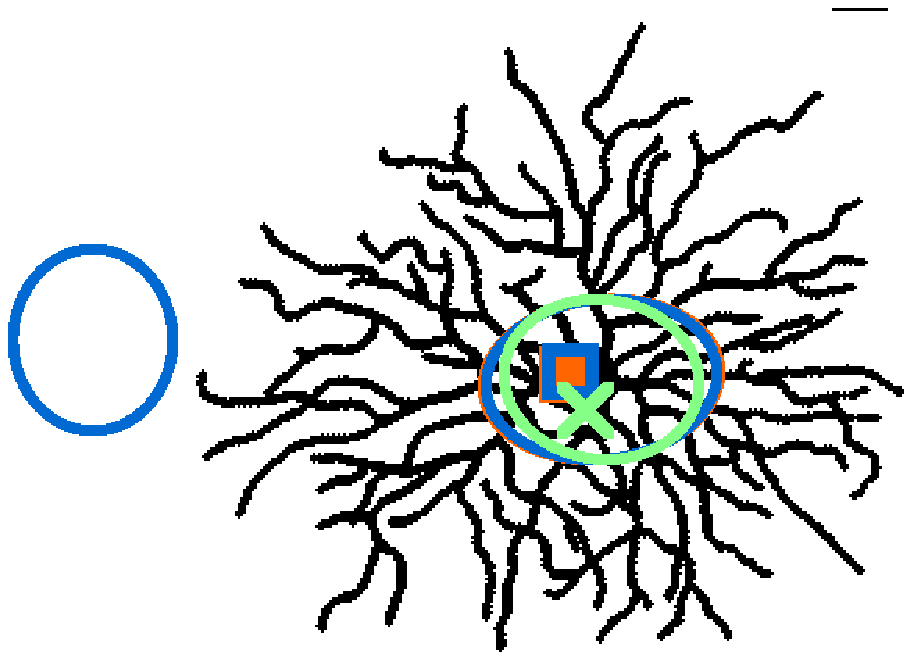}
\end{minipage}
\caption{Neurons of type $\alpha$  (top panel), $\beta$ (middle panel)
  and $\delta$ (bottom panel). The smoothing length is one pixel.  The
  meaning   of  the   points  and   the  ellipses   is   explained  in
  Fig.~\ref{fig:toy}. The small dash in the upper right corner of each
  panel has  a length of  20 pixels.  The  pictures are based  on data
  obtained  by~\cite{Masland:2001}  (their  figure no.   5,  copyright
  permission        by       \href{http://www.nature.com/neuro}{Nature
  Neuroscience}). Note, that  in all panels of this  figure as well as
  of  Figs.~\ref{fig:n2}  and  \ref{fig:n3}  the tensor  ellipses  for
  $V_0^{2,0}$ and $V_1^{2,0}$ almost coincide. }
\label{fig:n1}
\end{figure}
\begin{figure}
\centering\includegraphics[scale=0.55]{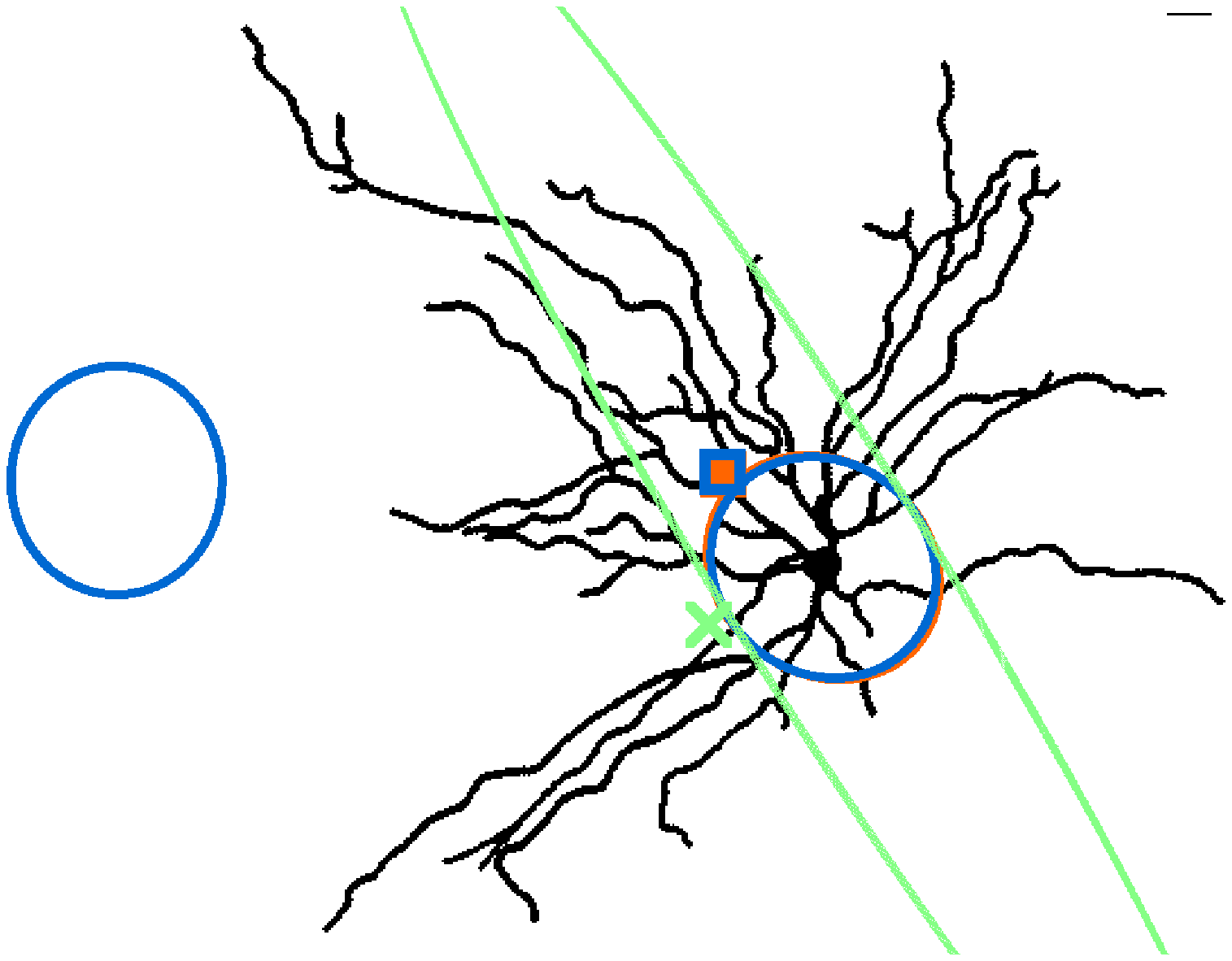}
\centering\includegraphics[scale=0.55]{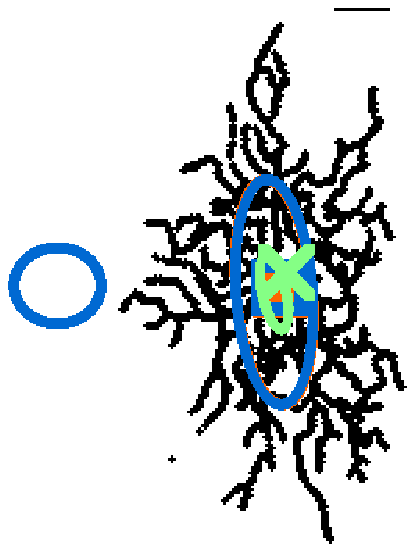}
\centering\includegraphics[scale=0.55]{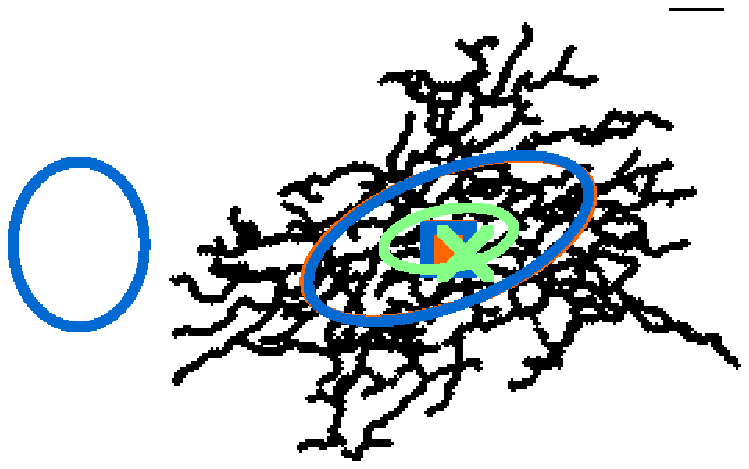}
\caption{Neurons of type $\epsilon$ (top panel), $\eta$ (bottom left
  panel) and $\iota$ (bottom right panel). The
  smoothing length
  is one pixel. The pictures are
  based  on  data obtained by~\cite{Masland:2001}  (their  figure  no.  5,
  copyright                        permission                       by
  \href{http://www.nature.com/neuro}{Nature Neuroscience}).  }
\label{fig:n2}
\end{figure}
\begin{figure}
\begin{minipage}[h]{.66\linewidth}
\centering\includegraphics[scale=0.55]{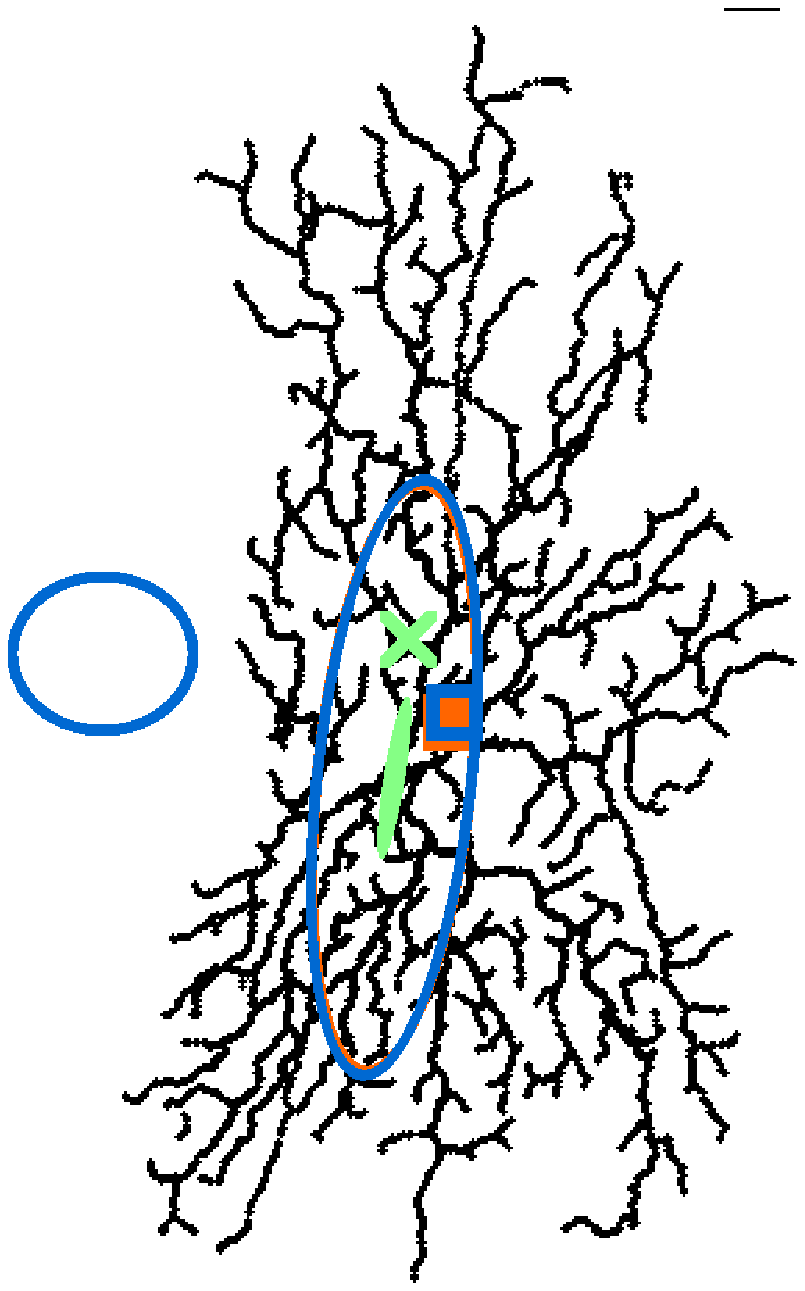}
\end{minipage}
\begin{minipage}[h]{.33\linewidth}
\centering\includegraphics[scale=0.55]{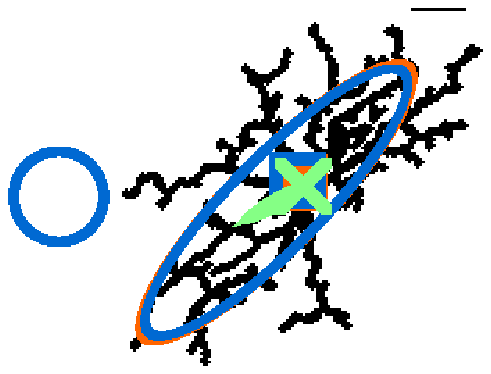}
\end{minipage}
\centering\includegraphics[scale=0.55]{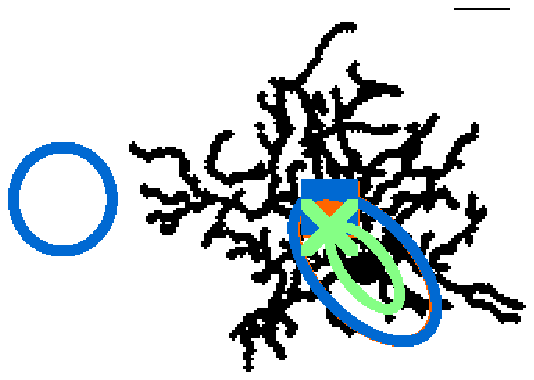}
\centering\includegraphics[scale=0.55]{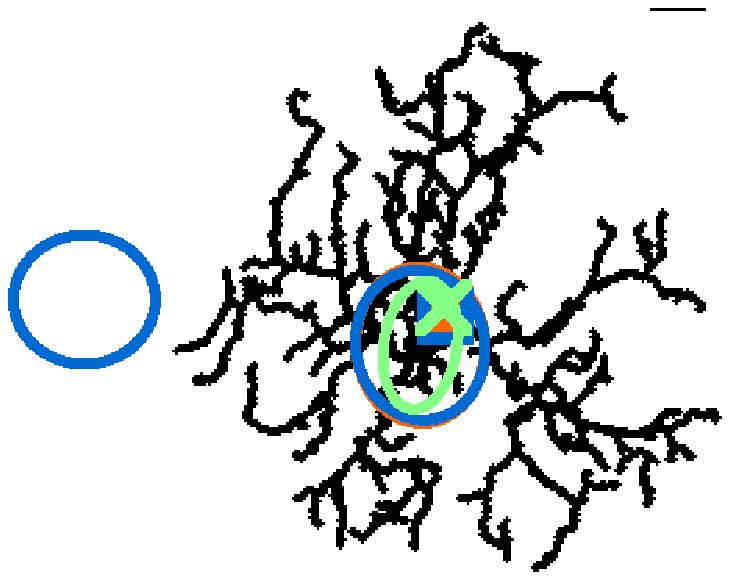}
\caption{Neurons of  type $\kappa$ (top panel), $\zeta$ (middle left
  panel), $\theta$ (middle right panel) and $\lambda$ (bottom 
  panel). The  smoothing length is  one pixel.  The pictures are
  based  on  data obtained by~\cite{Masland:2001}  (their  figure  no.  5,
  copyright                        permission                       by
  \href{http://www.nature.com/neuro}{Nature Neuroscience}).}
\label{fig:n3}
\end{figure}
\noindent 
\paragraph{Results.}
We show the neurons with some of  the results for a smoothing length of one pixel in Figs.~\ref{fig:n1}
-- \ref{fig:n3}.\footnote{ In the following, one has to be cautious in
  interpreting the green (light grey) ellipses,  because for our neuronal cells, the
  tensor $V_2^{2,0}$ sometimes has one or two negative eigenvalues. In
  this  case, the  ellipse  will  become smaller  and  point into  the
  direction of $\e_<$ instead of $\e_>$, if $|\tau_>|<|\tau_<|$. }
\\
Let us start with some qualitative observations.  First, the 
 centroids $\p_0$  through $\p_2$ are  typically not within  the soma.
Recalling  that   the centroids  are
 morphological centers, we can equivalently say that  the soma is quite often  eccentric.  It would
 be  interesting to know whether  the  eccentricity of  the soma  is
 characteristic for some  types of neurons (for this  we would have to
 investigate  larger statistical  samples of  neurons). We suspect
 that the eccentricities depend on the function and the local
 environment of the cells. Further investigations are needed to
 explore this effect.
\\
 Second,  we observe  that typically  $\p_0$ and  $\p_1$  almost coincide,
 whereas $\p_2$  may be further away  from them.  Something similar  is true
 about the tensors: The tensor ellipses of $V_0^{2,0}$ and $V_1^{2,0}$
 often closely resemble each other,  whereas  the ellipse  for $V_2^{2,0}$  greatly
 differs.  The  reason is as follows:  As our toy  examples have shown,
 $\p_2$,  $V_2^{2,0}$  and   the  corresponding  Minkowski  functional
 (viz. the Euler characteristic) are sensitive to holes.  For positive
 Euler characteristics, every hole that is off-soma pushes $\p_2$ onto
 the other side of the soma.  As a consequence, the location of $\p_2$
 and  the form of  $V_2^{2,0}$ very  much depend  on the  holes, their
 forms and positions. The holes in  turn depend on tiny details of the
 branching structure that  are not reflected in $\p_0$  and $\p_1$ and
 the corresponding tensors $V_0^{2,0}$  and $V_1^{2,0}$. -- Note, that
 most of  the holes are probably  due to the projection  of the neuron
 into two dimensions.
\\
We will now turn to a more quantitative analysis. We will show several
morphological  characteristics  that  are  based  upon  the  Minkowski
valuations as a function of  smoothing length $r_s$.  The point styles
designating the different kinds of  neurons are explained in the top
panel of Fig.~\ref{fig:smooth_sc0}.
\\
We show the first scalar  Minkowski functional $V_0$ for a large range
 of    smoothing   lengths    $r_s$    in   the  bottom   panel    of
 Figure~\ref{fig:smooth_sc0}.  For very small  $r_s$, $V_0$  grows very
 quickly, as $r_s$ increases;  whereas for larger smoothing lengths, a
 more moderate growth  can be seen. For some neurons  it appears to be
 linear,  for other  cell  types the  function  $V_0(r_s)$ is  clearly
 convex in this range. Bigger neurons typically grow faster than
 smaller ones.
 The  explanation is as follows: Let us consider
 the $\beta$ cell  first. Its overall shape is  roughly spherical, and
 its  extension $2r_0$ is  about $50$  pixels. If  the $\beta$  cell is
 smoothed with a very large $r_s>r_0=25$, all of its substructure is
 washed out, and we have approximately  the same result as if a circle
 of  radius  $r_0$ was  smoothed  by $r_s$.  So  the  volume is  about
 $V_0\approx \pi (r_0+r_s)^2 = \pi  r_0^2 + 2 \pi r_0 r_s+\pi r_s^2$,
 which is parabolic  in $r_s$. For $r_s<r_0$, the linear  term $ 2 \pi
 r_0 r_s$ is most significant, so the function $V_0(r_s)$ appears to
 be linear in a certain range. 
\\
More generally,  let $CP$ denote the  convex hull of a  pixelized data set
$P$ (or, more precisely, the pixel approximation of its convex hull). For large smoothing lengths, the parallel bodies of $P$ and $CP$,
$P_{r_s}$ and  $CP_{r_s}$ are very  close to each  other; consequently
the difference  $V_0\left(P_{r_s}\right)- V_1\left(CP_{r_s}\right)$ is
small compared to $V_0\left(P_{r_s}\right)$.  The size of
the parallel  body $CP_{r_s}$ can  be calculated using  {\em Steiner's
formula}     (see    \cite{weil:stereology},     p.     367,    e.g.):
\begin{equation}
V_0(CP_{r_s})=V_0(CP)+r_s  4  V_1  (CP)  + \pi  r_s^2\,.
\end{equation}
   This  again
defines a parabola, where the Minkowski functionals $V_0$ and $V_1$ of
$CP$  arise as  coefficients.  As  a  consequence, if  $r_s$ is  large
enough,  the  volume  $V_0(P_{r_s})$  is  largely  determined  by  the
Minkowski  functionals of the  convex hull  $CP$.  For  small neuronal
cells  such  as the  $\beta$  neuron,  this  behavior sets  in  quite
early. Bigger  neurons will have  larger values of  $V_0(CP)$ and  $V_1(CP)$ such
that their area $V_0$ is larger.\footnote{
Similar considerations apply to $V_1$ and $V_2$. 
} 
\\
In order to observe the fine-grained structure of the cells where the
neurons significantly differ from their convex hull, we have to
concentrate on smaller smoothing lengths $r_s<20$.  In
Figure~\ref{fig:smooth_sc} the scalar Minkowski functionals are
plotted vs.  the smoothing length $r_s$. For most neurons,  initially, $V_0$ grows comparatively
quickly; around $r_s=5$, however, the growth slows down.  As a
reason, the arms of the neurons that have been blown up, when
the parallel set was constructed, start to overlap with each other, such
that increasing $r_s$ will not necessarily fill many  pixels that have
not yet been occupied so far.
\\
For some bigger neurons ($\alpha$, $\delta$, $\kappa$, e.g.) a kind of
crossover  can be  observed around  $r_s=5$.  For  the other  types of
neuronal  cells, the crossover  is less  pronounced.\footnote{Note, by
the way,  that there  are plateaus  at the zero  points for  the $V_i$
vs. $r_s$  curves.  More generally,  these curves are  not continuous,
but change stepwise  because of our pixelwise smoothing.   This can be
seen, if  the $r_s$ resolution is  enhanced. In the  following we will
neglect discontinuities of this kind; they are a pure artifact of our
smoothing and do not carry any physical meaning. }
\\
\noindent We will now consider $V_1$. For small $r_s$, $V_1$
decreases as a function of $r_s$, since
 $V_1$ is dominated by small scale features that are smoothed
away stepwise. $V_1$ reaches a constant value later on.  This is  not what one  would expect  for a convex  body. The
reason, of  course, is that  the figure is  far from being  convex: As
$r_s$ increases, $V_1$ will gain at  the outer parts of the cells, but
loose in  the inner parts, because  holes are being  filled. Gains and
losses roughly compensate each other.   Note, that  the curves for the
$\alpha$, $\delta$ and  $\kappa$ cell type show an inflection point, which
very roughly  coincides with the position of their
crossover in $V_0$.
\\
\noindent The curves for the  Euler characteristic display a number of
discontinuities, but there is  some more general pattern. The negative
values indicate that the cells are dominated by holes.  For the bigger
cell types  ($\alpha$, $\delta$, $\epsilon$  and $\kappa$) there  is a
characteristic dip for small $r_s$. Up to this point, additional holes
are formed, as  branches of the neuron start to  touch each other. The
minimum  of the dip  roughly seems  to coincide  with the  point where
$V_0$ shows the crossover for the bigger neurons.  Similar results for
the Minkowski functionals have been obtained in~\cite{Barbosa:2003a}.
\\
A useful way of combining the information present in the scalar
Minkowski functionals is to construct the following dimensionless
quantity $Q$: 
\begin{equation}
Q := \frac{4 V_1^2}{\pi V_0} \, .
\end{equation}
This is a variation of the so-called isoperimetric ratio. For a convex
body  $P$  we have  $Q  (P)\ge  1$, where  the  equality  holds for  a
circle~\cite{fenchel:iso,alexandrov:iso,schmalzing:webI}.    $Q$    is
considerably larger  than one,  whenever the body  under investigation
has  an ``excess perimeter''  as compared  to its  area.  We  show the
logarithm of $Q$ as a function of $r_s$ in Fig.~\ref{fig:iso}.
The interpretation  is as follows: For small  smoothing lengths $r_s$,
most  of the  dendrites are  still present;  they produce  huge excess
areas for  which reason $Q$ starts  with very high  $Q$-values. As the
smoothing length  increases, $Q$ goes down.   The $\alpha$, $\epsilon$
and  $\kappa$   cells  have   the  largest  $Q$-values,   whereas  the
$\beta$-cell has the lowest $Q$-values  for a large range of smoothing
lengths because of its smallness  and its overall spherical shape. For
$r_s<6$  the  decrease  in  $\log_{10}(Q)$  seems roughly to  be
linear, where the slopes vary with the cell type.\footnote{
Note, that ``linearity'' holds only up to discreteness effects due to
our pixelwise smoothing.}
\\
In  Figs.~\ref{fig:smooth_dis} and  \ref{fig:smooth_dis2}  we consider
the  centroid distances  $\p_i$-soma,  $\dis_i$.  For  $i=0$ they  are
relatively  stable as  a function  of  $r_s$, whereas  for $i=1$  more
variation can be  observed. How is this to be  explained?  Look at the
$\kappa$ neuron  as an example (Fig.~\ref{fig:n3}). In  the lower half
of the  image the distribution of small  arms is a bit  denser than in
the upper half. Consequently, for  small $r_s$, there is a significant
contribution  to  the perimeter  from  this  part,  and this  is  also
reflected  in the  position  of $\p_1$,  which  is the  center of  the
perimeter. For larger $r_s\approx 10$, however, the lower, denser part
is filled more quickly, whereas in  the upper part quite big holes are
left, which  contribute to  the perimeter, such  that the  position of
$\p_1$ moves upwards. In this  way the curve for $\dis_1$ contains very
detailed information about the morphology of the neuron.
\\
In terms  of $\dis_0$ and  $\dis_1$ the soma  is most eccentric  for the
$\epsilon$  neuron.    This  is  also  reflected  in   our  visual
impressions.  It  might be useful,  however, to normalize  the $\dis_i$
parameters  by some  estimate of  the cell  size. If  we would  do so,
smaller  cells   would  have  a  reasonable  chance   of  having  bigger
eccentricities. 
\\
For $i=2$ (Fig.~\ref{fig:smooth_dis2}) we observe even larger
variations of the centroid distances.  Plateaus alternate with jumps
that can ultimately be traced back to discontinuities of the Euler
characteristic.  For small neurons, such as the $\beta$ type, however,
there is not much variation, because the cell is very small and gets
completely filled pretty soon.  For the $\alpha$, $\delta$ and
$\epsilon$-type, there is a common pattern:  As the smoothing length
increases, the jumps become larger.  The reason is probably, that for
larger smoothing lengths only a few holes will appear far off
centered.  When one of these outer holes disappears, $\p_2$ jumps
considerably.
\\
\noindent
In  Figures~\ref{fig:smooth_anis}  through  \ref{fig:smooth_anis5}  we
consider the anisotropy of the  cells. In order to quantify anisotropy
we  take the  eigenvalues  of the  tensors  $V_i^{j,k}$, $\tau_>$  and
$\tau_<$     and     calculate     the    quantity     $     \anis:=2(
\tau_>-\tau_<)/(|\tau_>|+|\tau_<|)\leq 2$.   The anisotropy parameters
derived from different tensors  focus on different kinds of anisotropy
(the  area   elements  belonging  to  a  body   might  be  distributed
differently from  those of its perimeter elements,  for instance).  As
can  be  seen  from  Fig.~\ref{fig:smooth_anis}, the  anisotropies  in
$V_0^{2,0}$ and $V_1^{2,0}$ are quite stable; most often they decrease
slowly, as
the smoothing length increases.  This indicates that the cells display
 large-scale  anisotropies that  are  not destroyed  by smoothing  the
 cell.  For  some cells  ($\eta$, $\kappa$, $\zeta$)  the anisotropies
 are considerable.   For each cell  type the anisotropies of  area and
 perimeter do not differ greatly. The $V_1^{2,0}$ tensor is a bit more
 sensitive  to  small-scale  variations  of  the  morphology, however;  so  the
 $\anis(V_1^{2,0})$-$r_s$   curves   appear   less  smooth   than   the
 $\anis(V_0^{2,0})$-$r_s$ curves.  On the  other hand, across the range
 of cell types, the variation is quite high. Thus anisotropies seem to
 have a significant discriminative power.
\\
It is  different with the  tensor $V_2^{2,0}$, which is  considered in
  Figure~\ref{fig:smooth_anis4}.   The  anisotropy  derived from  this
  tensor jumps back and forth and sometimes reaches values that exceed
  those derived  from the other  tensors. This performance  should not
  come  as  a  surprise,  since   we  have  already  seen  that  other
  characteristics  that  are   related  to  the  Euler  characteristic
  such as  $V_2^{2,0}$
  typically  show discontinuities.  At some  point, however,  when the
  smoothing has produced one  connected pattern without holes (visible
  for  the $\beta$  cell,  e.g.,  where this  point  is reached  quite
  early), the  anisotropy stabilizes at a constant  value.  Apart from
  this, the dependence  on $r_s$ looks rather chaotic;  so far, we are
  not  able to  extract information  that might  help  to discriminate
  between the different cell types.
\\
As  mentioned before, on  the square  lattice, the  last tensor  to be
considered,  $V_1^{0,2}$,  has  a  simple interpretation.   It  checks
whether the majority  of normals are parallel to  the horizontal or to
the vertical  grid axis. If $\partial  P$ is dominated  by vertical or
horizontal   normals,  $V_1^{0,2}$   will   display  a   corresponding
anisotropy;  if  not,  $V_1^{0,2}$  will  roughly  be  isotropic.   In
Figure~\ref{fig:smooth_anis5} we show some results for single neurons.
One can learn from this that the anisotropies arising from $V_1^{0,2}$
are quite small. The anisotropy  is comparatively large for the $\eta$
type cell, because this cell is clearly elongated. For small values of
the smoothing length, $\anis(V_1^{0,2})$  is not so much influenced by
the overall shape  of the neuron, but rather by  the directions of the
single  arms.   Interestingly,  the  graphs  shown  are  qualitatively
different  for  the different  types  of  cells:  One cell  (viz.  the
$\alpha$ cell) starts with  zero anisotropy, whereas others begin with
a   non-zero  anisotropy.   Moreover,   there  are   significant  peak
structures.  But because  of its relation to normals,  $\n$, the value
of $V_1^{0,2}$  depends to  a large extent  on the orientation  of the
cell with respect to the grid.  For this reason $V_1^{0,2}$ is only of
limited use.
\\
In   Fig.~\ref{fig:smooth_tr},  the  {\em   traces}  of   the  tensors
$V_i^{2,0}$  are   considered  (the   trace  of  the   fourth  tensor,
$V_1^{0,2}$ need not  to be taken into account  at this point, because
it  equals  $V_1$).  Qualitatively,  the  viewgraphs  for  $V_i^{2,0}$
resemble the curves
 of  their scalar  counterparts, $V_i$  for $i=0,..,2$.   In  order to
 extract more specific  information, it is thus useful  to divide $\tr
 \left(V_i^{2,0}\right)$  by $V_i$,  respectively, for  $i=0,..,2$. The  result is a measure of
how concentrated a cell is in terms of area,  perimeter or curvature: $\tr
(V_0^{2,0})/V_0$, for instance will be  the bigger, the further the soma
and those  parts of the cell that  bear most of its  volume lie apart.  
Results   can   be   seen  from   Figure~\ref{fig:smooth_trn}.   $\tr ((V_0^{2,0})/V_0$  increases continuously, as $r_s$
 is enhanced. The reason is that more and more pixels are added at the
 outer  parts of  the  neuron, so  the  neuron becomes  less and  less
 concentrated. In $\tr ((V_1^{2,0})/V_1$ there is a kink at
 least for some neurons ($\alpha$, $\delta$, $\epsilon$, $\kappa$). It
 indicates an additional growth  effect. Very probably the explanation
 is  that  for  small  $r_s$,  the  small  branches  within  the  cell
 significantly contribute to $V_1^{2,0}$,  so the neuron appears to be
 very concentrated; for  larger values of $r_s$ the  arms merge and do
 not contribute to the perimeter any more,  so most of the neuron's perimeter
 is found at  its outer parts. Note, that the kinks  roughly set in at
 the  $r_s$-locations  of  the   crossover  point  in  $V_0$  and  the
 inflection  point in $V_1$  for the  $\alpha$, $\delta$  and $\kappa$
 cells.

\begin{figure}
\centering\includegraphics[width=8.4cm]{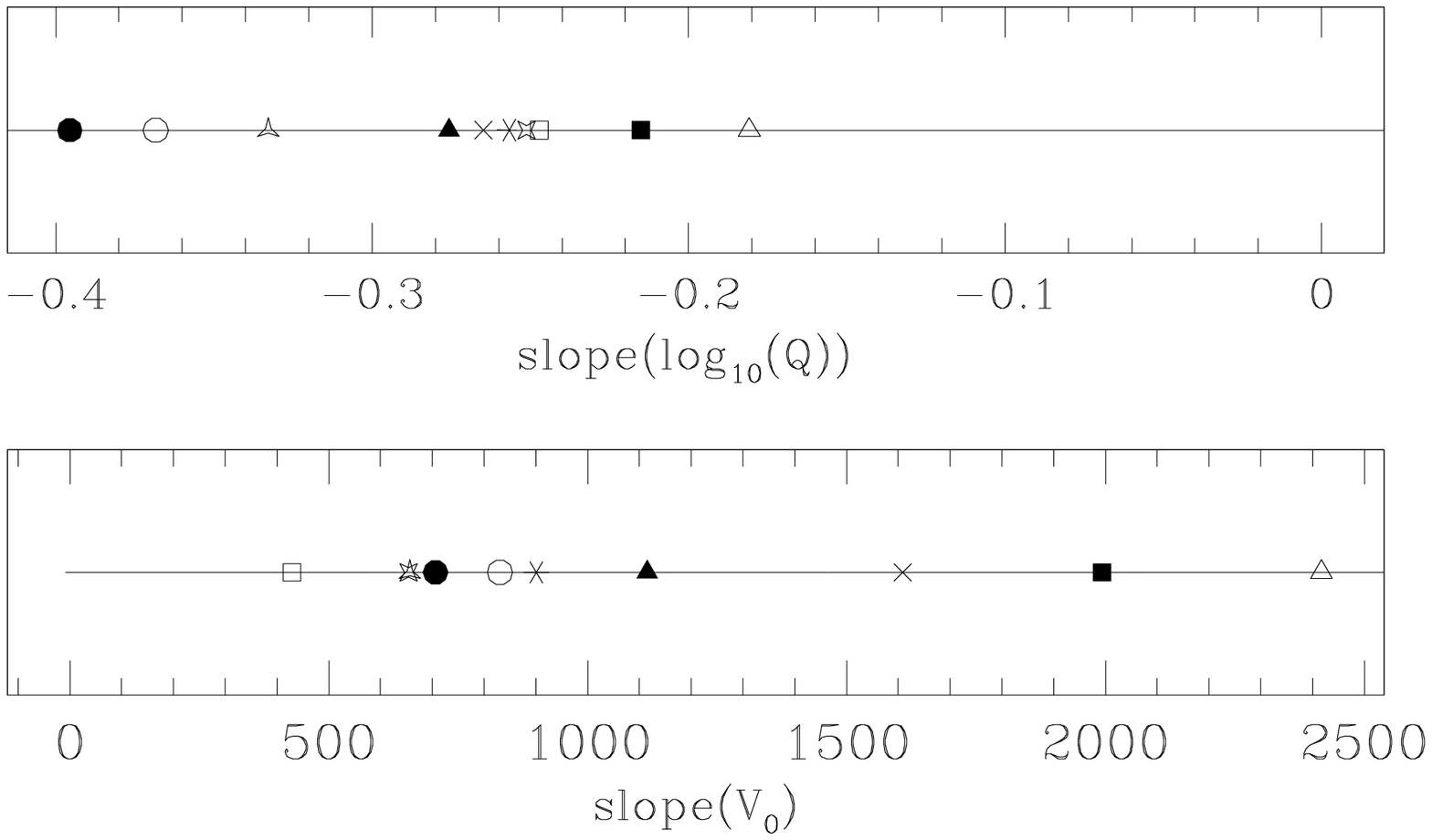}
\caption{The slopes of $V_0$ (top panel) and $\log_{10}(Q)$ for the
  different cells. The point styles are as above. }\label{fig:slopes}
\end{figure}

\begin{figure}
\begin{center}
\subfigure[]{\includegraphics[width=7cm]{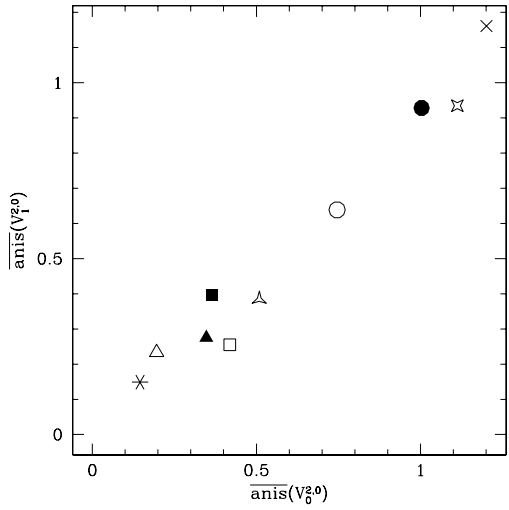}}
\subfigure[]{\includegraphics[width=7cm]{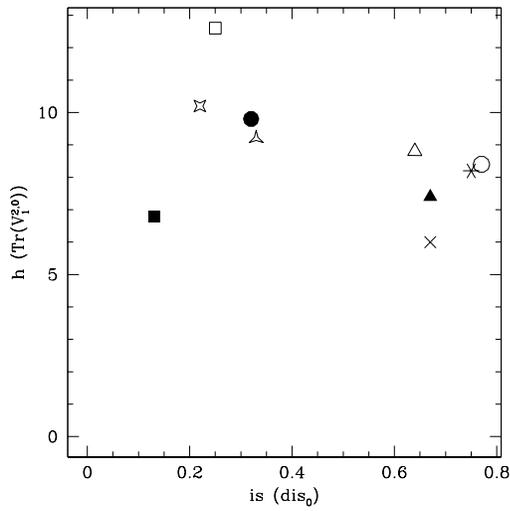}}
\caption{Scatter  plots  from  selected  features of  an extended  Minkowski
  analysis  showing  the population  of  the  feature  space with  the
  neuronal cells.~\label{fig:scatter}}
\end{center}
\end{figure}

\begin{figure}
\centering\includegraphics[width=11.2cm, angle=270]{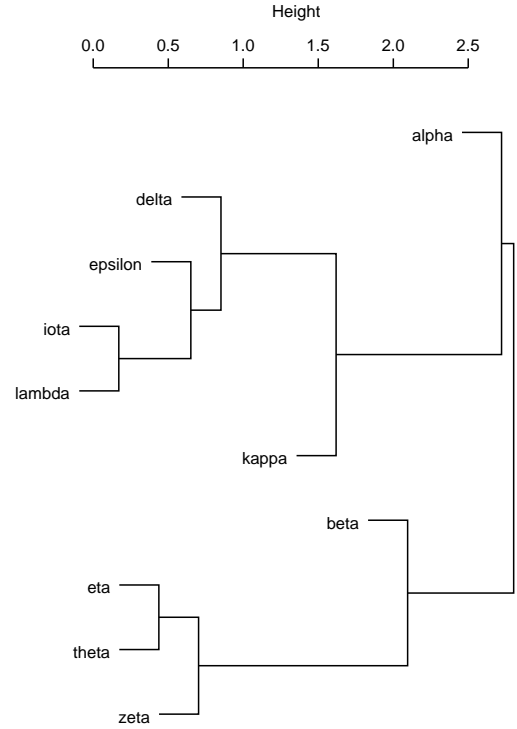}
\caption{
  The  classification  pattern  according to   an  agglomerative  hierarchical
  clustering analysis  considering the  two features selected  for the
  scatter plot in Figure~\ref{fig:scatter}(b).~\label{fig:tree}}
\end{figure}

\paragraph{Constructing global measures}
A multiscale analysis  like that presented in this  paper,  leads
to  rich and  detailed information  on the  geometrical aspects  of an
object.  Nevertheless,  once such a  description of the data  has been
obtained,  it is  often useful  to derive  a compact  set  of global
measures that summarize the most important morphological aspects. In this
paper, we consider several ways of condensing multiscale information,
i.e. a function of some scale, into
simple       parameters:        The  \emph{monotonicity
index}~\cite{Barbosa:2003a,Barbosa:2003b}   is defined as
\begin{equation}
i_s=\frac{s}{s+d+p}\,,
\end{equation}
where  $s$,  $d$ and  $p$  count  each  time the  function  increases,
decreases and remains  unchanged, respectively.  Thus $i_s$ quantifies
the  fraction of  the  interval where  the  function is  monotonically
increasing. The \emph{mean value} is the average value of the function
over the  interval.  The \emph{half scale}  is the scale  at which the
area below a curve reaches half of its total value. A different way of
constructing  global  parameters is  to  consider  the  \emph{slope} of  some
characteristic in some particular range of $r_s$-values.
\\
In Figure~\ref{fig:slopes} we visualize the average slopes of $V_0$ in the range
$r_s\in [10,60]$ and of  $\log_{10}(Q)$ in the range $r_s\in[1.5,4.5]$
for  the  different  cells.  In  both  cases  we  choose  a  range  of
$r_s$-values  for  which  the  functionals  under  investigation  look
roughly   linear  for   most  cell   types.  Results   are   shown  in
Figure~\ref{fig:slopes}.  One  can  immediately  see that  the  slopes
discriminate amongst the different cell types.
\\
In order  to further illustrate our approach, we  selected two two-dimensional
feature  spaces, which are  spanned by  size-independent morphological
characteristics.   In  order  to  calculate them,  we  considered  the
interval $r_s \in \left[0,20\right]$ and a spacing of $0.2$.
\\
Our first feature space is spanned by the mean of the anisotropy parameter
derived from $V_0^{2,0}$, $\overline{\anis}\left(V_0^{2,0}\right)$, and the
mean of the anisotropy parameter corresponding to $V_0^{2,0}$, $\overline{
  \anis}\left(V_1^{2,0}\right)$.  It is shown in Figure~\ref{fig:scatter}(a).
There appears to be some systematic correlation between both characteristics:
cells with higher $\overline{ \anis}\left(V_0^{2,0}\right)$ tend to have
higher $\overline{ \anis}\left(V_1^{2,0}\right)$ as well. Given the meaning of
these characteristics, this should not come as a surprise, although it is in
principle possible to have high anisotropy in $V_0^{2,0}$ and low anisotropy in
$V_1^{2,0}$.  Thus, for discriminating between different cells, one dimension
of this feature space is essentially redundant. But the presence of some
correlation might be used to describe some common trait shared by all cells.
\\
A different situation can be observed for our second feature space. It is spanned
by the monotonicity index $i_s\left(\dis_0\right)$ and by the half scale
$h\left(\tr\left(V_1^{2,0}\right)\right)$.  As can be seen from
Figure~\ref{fig:scatter}(b), the scatter is larger, and the cells form kind of
groups.  Note, in particular, that neuronal cells that look similar at least in
some respect tend to appear close to each other in this scatter plot.  For
instance, cells $\beta$, $\eta$, $\theta$ and $\zeta$ are close to each other
in the bottom panel of Figure~\ref{fig:scatter}, especially regarding the
position of their center of mass $\\p_0$ relatively to their soma.
Figure~\ref{fig:tree} presents a dendrogram obtained by a simple hierarchical
agglomerative clustering~\cite{Cluster:book} of the scatter plot distribution
shown in Figure~\ref{fig:scatter}(b).  Such a structure suggests a possible
\emph{taxonomy} for the ten types of cells.  As expected, the cells $\beta$,
$\eta$, $\theta$ and $\zeta$ are similar, inhabiting the same branch at the
lower part of the dendrogram.  For the remaining subset, the cells
$\alpha$ and $\kappa$ end up markedly distinct from the group of cells formed
by $\delta$, $\epsilon$, $\iota$ and $\lambda$.
\\
Although the proposed  methodology may have a bearing on 
the classification of cat  ganglion cells, it is difficult to make
more 
definitive        conclusions at this point,        because       the        original
classification~\cite{Berson:2002}  takes  into  account not  only  the
neuronal morphology, but also the  cell stratification and the size of
the soma.  Moreover,  except for the more common  $\alpha$ and $\beta$
types, only a small number of  examples of the cell types have been
analyzed in the related literature~\cite{Berson:2002}. A more
detailed examination of which feature spaces are most useful has
to wait for further data.
%%%%%%%%%%%%%%%%%%%%%%%%%%%%%%%%%%%%%%%%%%%%%%%%%%
\section{Conclusions}  
%%%%%%%%%%%%%%%%%%%%%%%%%%%%%%%%%%%%%%%%%%%%%%%%%%

We have  analyzed two-dimensional projections of  neuronal cells using
higher-order Minkowski valuations. Our measures detect different kinds
of substructures, providing a natural extension of previous works that
deal        with        the        more       traditional        shape
functionals~\cite{Barbosa:2003a,Barbosa:2003b}.       An     extensive
discussion of the results obtained for a set of ten neuronal cells was
included that illustrates the interpretation of the suggested measures
and implications for neuromorphometric studies.  As far as our limited
set of samples is  concerned, significant similarities and differences
between  the  cell  types  have  been found,  leading  to  a  putative
taxonomy.  It is a pending question whether the differences found will
still be  characteristic of the  types in a  statistical sense.\\[2cm]
{\small  This  work was  financially  supported  by FAPESP  (processes
  02/02504-01 and 99/12765-2) and  CNPq (process 308231/03-1).  It was
  also   supported   by   the  "Sonderforschungsbereich   375-95   für
  Astro-Teilchenphysik"  der Deutschen  Forschungsgemeinschaft.   C.B. 
  thanks  the Alexander  von Humboldt  Foundation, the  German Federal
  Ministry  of  Education  and   Research  and  the  Program  for  the
  Investment  in  the  Future  (ZIP)  of  the  German  Government  for
  supporting  this  research.  He  also  thanks  Jens  Schmalzing  for
  providing software  on which  parts of the  codes for this  paper are
  built upon.}

\bibliographystyle{vunsrt}
\bibliography{n}

\end{document}